\documentstyle[preprint,tighten,amsfonts,eqsecnum,aps,prd]{revtex}

\begin{document}
\draft

\hyphenation{
mani-fold
mani-folds
geo-metry
geo-met-ric
}



\def\BbbR{{\Bbb R}}
\def\BbbZ{{\Bbb Z}}
\def\BbbC{{\Bbb C}}

\def\half{{\frac{1}{2}}}
\def\casehalf{{\case{1}{2}}}

\def\SLtwor{{\rm SL}(2,\BbbR)}
\def\sltwor{{\frak sl}(2,\BbbR)}

\def\SLtwordouble{{\widetilde{\rm SL}}(2,\BbbR)}
\def\PSLtwor{{\rm PSL}(2,\BbbR)}

\def\Ogrouprs{{\rm O}(r,s)}

\def\Otwotwo{{\rm O}(2,2)}
\def\Octwotwo{{\rm O}_{\rm c}(2,2)}
\def\otwotwo{{\frak o}(2,2)}

\def\Otwoone{{\rm O}(2,1)}
\def\Octwoone{{\rm O}_{\rm c}(2,1)}

\def\Aclass{{\cal A}_{\rm class}}

\def\Aphy{{\cal A}_{\rm phy}}
\def\Aphystar{{\cal A}_{\rm phy}^{(\star)}}
\def\Aphyplusstar{{\cal A}_{{\rm phy}+}^{(\star)}}

\def\Haux{{\cal H}_{\rm aux}}
\def\Hphys{{\cal H}_{\rm phys}}

\def\Aobs{{\cal A}_{\rm obs}}
\def\Aobsplus{{\cal A}_{\rm obs}^+}

\def\Agrouphat{{\hat{\cal A}}_{G}}


\preprint{\vbox{\baselineskip=12pt
\rightline{AEI 1999-010}
\rightline{gr-qc/9907004}}}

\title{Refined Algebraic Quantization in the 
\\
oscillator representation 
of $\SLtwor$}
\author{Jorma Louko\footnote{%
Electronic address:
jorma.louko@nottingham.ac.uk. 
Address after September~1, 1999: 
School of Mathematical Sciences, 
University of Nottingham, 
Nottingham NG7 2RD, U.K\null. 
}}
\address{
Max-Planck-Institut f\"ur Gravitations\-physik,
Am M\"uhlenberg~5, 
D-14476 Golm, 
Germany
}
\author{Carlo Rovelli\footnote{Electronic address:
rovelli@cpt.univ-mrs.fr}}
\address{
Department of Physics and Astronomy, 
University of Pittsburgh, 
\\
Pittsburgh, Pennsylvania 15260, USA
\\
and
\\
Centre de Physique Theorique, 
CNRS Luminy, 
F-13288 Marseille, 
France
}
\date{Journal of Mathematical Physics {\bf 41} (2000) 132--155}
\maketitle 
\begin{abstract}%
We investigate Refined Algebraic Quantization (RAQ) with group
averaging in a constrained Hamiltonian system with unreduced phase
space $T^*\BbbR^4$ and gauge group~$\SLtwor$.  The reduced phase
space ${\cal M}$ is connected and contains four mutually
disconnected `regular' sectors with topology $\BbbR\times S^1$, but
these sectors are connected to each other through an exceptional set
where ${\cal M}$ is not a manifold and where ${\cal M}$ has
non-Hausdorff topology.  The RAQ physical Hilbert space $\Hphys$
decomposes as $\Hphys \simeq \oplus {\cal H}_i$, where the four
subspaces ${\cal H}_i$ naturally correspond to the four regular
sectors of~${\cal M}$. The RAQ observable algebra~$\Aobs$,
represented on~$\Hphys$, contains natural subalgebras represented on
each ${\cal H}_i$. The group averaging takes place in the oscillator
representation of $\SLtwor$ on $L^2(\BbbR^{2,2})$, and ensuring
convergence requires a subtle choice for the test state space: the
classical analogue of this choice is to excise from ${\cal M}$ the
exceptional set while nevertheless retaining information about the
connections between the regular sectors.  A~quantum theory with the
Hilbert space $\Hphys$ and a finitely-generated observable
subalgebra of $\Aobs$ is recovered through both Ashtekar's Algebraic
Quantization and Isham's group theoretic quantization.
\end{abstract}

\narrowtext 


\section{Introduction}
\label{sec:intro}

In the quantization of constrained systems, one proposal for defining
an inner product on the physical Hilbert space is to induce this inner
product from an auxiliary Hilbert space $\Haux$ via averaging over the
gauge group.  The construction of $\Haux$ draws input from the
kinematical structure of the theory before imposing the constraints,
and the constraints enter through an operator representation of the
gauge group on~$\Haux$.  The method has emerged and been applied in
various contexts; see
\cite{AH,KL,QORD,epistle,BC,lands-against,lands-wren,giumar-unique,GoMa}
and the references therein.

A~major open question with group averaging is 
the sense in
which the averaging can be made to converge.
One may encounter situations where the group averaging diverges 
merely because of some ill-chosen piece of technical input, 
and modifying the input leads to a well-defined theory.  On the other
hand, one may also encounter situations where convergence of the group 
averaging is precluded by some physically interesting property of the
system. For example, within the Refined Algebraic Quantization 
framework of
\cite{giumar-unique}, a convergent group averaging cannot yield a
theory with superselection sectors, while a well-defined theory with
superselection sectors may nevertheless be recovered through a suitable
renormalization of the averaging \cite{GoMa}. 

In this paper we study group averaging in a quantum mechanical system
whose constraints generate the gauge group~$\SLtwor$.  The classical
phase space is $\Gamma=T^*\BbbR^4$, and the three classical
constraints on $\Gamma$ 
are homogeneous quadratic functions of the global canonical phase
space coordinates.  The system was introduced by Montesinos, Rovelli, 
and Thiemann \cite{monrovthie} 
as an analogue of general relativity with two
``Hamiltonian''-type constraints, quadratic in the momenta, and one
``momentum''-type constraint, linear in the momenta.  The reduced
phase space ${\cal M}$ is connected, and it contains four mutually
disconnected `regular' sectors with topology 
$\BbbR\times S^1$, but connecting
these sectors there is an exceptional set where ${\cal M}$ is not a
manifold and the topology of ${\cal M}$ is non-Hausdorff. 
One thus anticipates quantization to produce a theory with 
four `regular' sectors, with subtleties 
in those aspects of
quantization that try in some sense 
to connect these four sectors. 
We shall see that this 
is indeed the case, and when group averaging is used in the
quantization, 
the subtleties emerge precisely in the
convergence of the group averaging. 

We consider two quantization approaches. 
First, we recall that $\Gamma$ admits an explicitly-known 
$\otwotwo$ Poisson bracket 
algebra $\Aclass$ of constants 
of motion (``observables'') that separates the regular sectors 
of ${\cal M}$ \cite{monrovthie}. 
We therefore carry through Ashtekar's Algebraic Quantization program 
\cite{AAbook,ashtate} 
with
$\Aclass$ promoted into a quantum observable star-algebra~$\Aphystar$. 
In agreement with the results of \cite{monrovthie}, we find
four distinct Hilbert spaces, each corresponding to 
one of the
regular sectors of~${\cal M}$.
We then add to $\Aphystar$ four new generators whose classical
counterparts act on the four sectors of ${\cal M}$ as a $\BbbZ_2\times 
\BbbZ_2$ permutation subgroup, and we 
carry through Algebraic Quantization with the resulting larger observable
algebra~$\Aphyplusstar$. 
Expectedly, the emerging Hilbert space ${\cal H}_+$ is the
direct sum of the previous four individual Hilbert spaces. 
We also show that ${\cal H}_+$ with the observable algebra 
$\Aphyplusstar$ can be recovered by applying Isham's group theoretic
quantization \cite{isham-les} 
to an $\Otwotwo$ action on~$\Gamma$: the
infinitesimal generators of the action of the connected subgroup 
$\Octwotwo$ are precisely the classical
observables in~$\Aclass$. 

We then consider a group averaging approach. 
For concreteness, and to
a considerable degree without loss of generality
\cite{giumar-general}, we adopt the formalism of Refined Algebraic
Quantization (RAQ) \cite{epistle,giumar-unique,giumar-general}.  The
structure of $\Gamma$ and the classical constraints suggests a natural
choice for $\Haux$ and for the representation of the gauge group
$\SLtwor$: this representation is isomorphic to the 
oscillator representation of $\SLtwor$ on $L^2(\BbbR^{2,2})$
\cite{howe-tan}.  $\Haux$~also carries 
a representation of the Algebraic
Quantization observable algebra~$\Aphyplusstar$, and this
representation commutes with the $\SLtwor$-action.  With a suitable
choice for the RAQ linear space 
$\Phi\subset\Haux$ of test states, we find
that the group averaging converges in absolute value and produces
a nontrivial physical Hilbert
space~$\Hphys$. 
$\Hphys$~is isomorphic
to~${\cal H}_+$, and the representation of the RAQ observable algebra
$\Aobs$ 
on $\Hphys$ contains a subrepresentation isomorphic to the
representation of $\Aphyplusstar$ on~${\cal H}_+$. (For technical
reasons, 
these isomorphisms are antilinear.) In this sense, the
RAQ quantum theory contains the Algebraic Quantization quantum theory.
Further, the uniqueness theorem of \cite{giumar-unique} shows that our
choices for~$\Haux$, the $\SLtwor$-action, and $\Phi$ completely
determine the RAQ quantum theory even
without group averaging: with our choices, 
the only freedom in the RAQ rigging
map is a multiplicative constant. 

Now to the promised subtleties. 
In the Algebraic Quantization approach, the subtlety occurs
with the 
choice of the linear space on which the constraints are solved. The
`natural' first candidate ${\tilde V}$ for this linear space contains
a one-dimensional subspace that, by the spectral properties
of~$\Aphystar$, corresponds classically to the exceptional set
in~${\cal M}$.  This subspace turns however out to have zero norm, and
one does not recover a Hilbert space. The remedy is simply to drop the
troublesome one-dimensional subspace from~${\tilde V}$,
with the results mentioned above.

In the RAQ approach, the subtlety occurs 
with the choice of the test
state space. 
The structure
of the quantum constraint operators and the $\SLtwor$-action 
suggests a natural choice~${\tilde \Phi}$, 
but it 
turns out that the group averaging fails to converge precisely
on the subspace of ${\tilde \Phi}$ 
where it attempts to produce the
``zero norm'' vectors encountered in the 
Algebraic Quantization. 
The remedy is again to ensure that the
troublesome subspace does not appear in the physical Hilbert space, 
but now this has to be done by modifying the test state space, and as
the definition of observables in RAQ is intimately related to the test 
state space, care must be taken in order that the RAQ
observable algebra remain large enough to allow a comparison with the
Algebraic Quantization observable algebra. 
Our choice, $\Phi$, was found by scrutinizing the explicitly-known 
$\Aphyplusstar$-action on~${\tilde \Phi}$. 

The rest of the paper is as follows. In section 
\ref{sec:classical} we review and analyze the classical system
\cite{monrovthie}, 
paying special attention to the classical 
observable algebra~$\Aclass$, 
its pull-backs to the various parts of the reduced phase
space, and the associated $\Otwotwo$ action. The Algebraic
Quantization and the group theoretic quantization are
carried out in section~\ref{sec:oldAQ}. Section 
\ref{sec:RAQoutline} presents a concise outline of RAQ with group
averaging, in the specific formulation of \cite{giumar-unique}, and
section \ref{sec:RAQus} 
carries out RAQ in our system. 
Section \ref{sec:discussion} contains a brief discussion. 
Appendices \ref{app:iwasawa} and \ref{app:oscrep} collect some 
relevant facts about 
$\SLtwor$, its covering groups, and their
oscillator representations \cite{howe-tan}. 
Certain technical calculations concerning the group averaging are 
given in appendices \ref{app:convergence} and~\ref{app:rigg-ev}.

\section{Classical dynamics}
\label{sec:classical}

In this section we 
review and 
analyze the classical system introduced in
\cite{monrovthie}. 
Some relevant facts about the group $\SLtwor$ and its Lie algebra 
$\sltwor$ are collected in
appendix~\ref{app:iwasawa}. 

The phase space is $\Gamma := T^*\BbbR^4\simeq \BbbR^8$, 
with the global coordinate
functions $(u^1,u^2,v^1,v^2)$ for the base and $(p^1,p^2,\pi^1,\pi^2)$
for the cotangent fibers. The symplectic structure is 
$\Omega = \sum_i \left( dp^i \wedge du^i +  d\pi^i \wedge dv^i
\right)$. 
We adopt the vector notation
$(u^1,u^2):=\vec u$, $(v^1,v^2):=\vec v$, $(p^1,p^2):=\vec p$,
$(\pi^1,\pi^2):=\vec \pi$, and we indicate a contraction in the
suppressed two-dimensional indices by a dot product. 

The action reads
\begin{equation}
S  = \int dt \left(
{\vec p} \cdot 
{\dot{\vec u}}
+ 
{\vec {\pi}} \cdot 
{\dot {\vec v}} 
 - N^1 H_1
 - N^2 H_2
 - \lambda D
 \right)
\ \ ,
\label{action}
\end{equation}
where $N^1$,  $N^2$, and $\lambda$ are Lagrange multipliers, and the
three constraints are 
\begin{mathletters}
\label{constraints}
\begin{eqnarray}
H_1 
& := & 
\casehalf
\left( {\vec p \, }^2 - \vec v^2 \right)
\ \ ,
\\
H_2 
& := & 
\casehalf
\left( \vec \pi^2 - \vec u^2 \right)
\ \ ,
\\
D & := & 
{\vec u}\cdot {\vec p} - 
{\vec v} \cdot {\vec \pi}
\ \ .
\end{eqnarray}
\end{mathletters}
The Poisson bracket algebra of the constraints is 
\begin{mathletters}
\label{PBalg-constraints}
\begin{eqnarray}
\{ H_1 \, , \, H_2 \} 
& = & 
D
\ \ ,
\\
\{ D \, , \, H_1 \} 
& = & 
2H_1
\ \ ,
\\
\{ D \, , \, H_2 \} 
& = & 
- 2H_2
\ \ ,
\end{eqnarray}
\end{mathletters}
which is isomorphic to the 
Lie algebra $\sltwor$ in the basis (\ref{sltwo-basis}) of 
appendix~\ref{app:iwasawa}. 
The system is therefore a first class constrained system
\cite{Henneaux}. 
The gauge group generated by the constraints is 
$\SLtwor$, and its action on $\Gamma$ is 
\cite{monrovthie}
\begin{eqnarray}
&&
\pmatrix{\vec u \cr \vec p \cr} 
\mapsto 
g \pmatrix{\vec u \cr \vec p \cr} 
\ \ ,
\nonumber
\\
&&
\pmatrix{\vec \pi \cr \vec v \cr} 
\mapsto 
g \pmatrix{\vec \pi \cr \vec v \cr} 
\ \ ,
\label{class-gaugeaction}
\end{eqnarray}
where  $g$ is an $2\times2$ matrix in $\SLtwor$. 

The reduced phase space ${\cal M}$ is, by definition, the 
quotient of the constraint hypersurface under the 
$\SLtwor$ action~(\ref{class-gaugeaction}). 
The topology of ${\cal M}$ is induced from~$\Gamma$, and wherever the 
geometry of ${\cal M}$ is sufficiently regular, 
${\cal M}$ inherits from $\Gamma$ also a differentiable structure and
a real analytic structure. 

${\cal M}_0$~decomposes
naturally into six subsets, which we denote
respectively by 
${\cal M}_0$, ${\cal M}_{\rm ex}$, 
and ${\cal M}_{\epsilon_1,\epsilon_2}$, where
$\epsilon_i\in\{1,-1\}$. 
For the points in~${\cal M}_{\epsilon_1,\epsilon_2}$, unique
representatives in $\Gamma$ are
\begin{eqnarray}
&&
\vec u = \sqrt{r} \, ( 1,0 )
\ \ ,
\nonumber
\\
&&
\vec p = \sqrt{r} \, ( 0 ,\epsilon_1 )
\ \ ,
\nonumber
\\
&&
\vec v = \sqrt{r} \,
( \cos \varphi, - \epsilon_1 \epsilon_2 \sin\varphi )
\ \ ,
\nonumber
\\
&&
\vec \pi = \sqrt{r} \,
( \sin \varphi, + \epsilon_1 \epsilon_2 \cos\varphi )
\ \ , 
\label{Mgen-reps}
\end{eqnarray}
where $r>0$ and $0\le\varphi<2\pi$. 
For the points in~${\cal M}_{\rm ex}$, unique representatives in
$\Gamma$ are 
\begin{eqnarray}
&&
\vec u = ( \cos\theta,\sin\theta )
\ \ ,
\nonumber
\\
&&
\vec \pi = ( \cos\varphi,\sin\varphi)
\ \ ,
\nonumber
\\
&&
\vec v = \vec p = 0 
\ \ ,
\label{Mex-reps}
\end{eqnarray}
where $0\le\theta<\pi$ and $0\le\varphi<2\pi$. 
${\cal M}_0$~contains a single
point, whose unique representative in $\Gamma$ is $\vec u = \vec v =
\vec p = \vec \pi = 0$. 

The four subsets ${\cal M}_{\epsilon_1,\epsilon_2}$ of ${\cal M}$ are 
disconnected. Each is open in ${\cal M}$ and 
has topology $\BbbR \times S^1$, and each is 
coordinatized by the pair $(r,\varphi)$ as shown in~(\ref{Mgen-reps}),
with $r>0$ and $(r,\varphi)\sim(r,\varphi+2\pi)$. The pullback of
$\Omega$ to each ${\cal M}_{\epsilon_1,\epsilon_2}$ is nondegenerate
and equal to $-dr \wedge d\varphi$, thus making each 
${\cal M}_{\epsilon_1,\epsilon_2}$ into a smooth 
symplectic manifold. 
We regard 
${\cal M}_{\epsilon_1,\epsilon_2}$ 
as the four `regular' sectors of~${\cal M}$, 
and we denote their union by ${\cal M}_{\rm reg}$. 

${\cal M}_{\rm ex}$ is a smooth two-dimensional manifold, and the 
pullback of $\Omega$ to ${\cal M}_{\rm ex}$ vanishes. The topology of 
${\cal M}$ near ${\cal M}_{\rm ex}$ is severely non-Hausdorff: 
any neighborhood of any point in 
${\cal M}_{\rm ex}$ contains~${\cal M}_0$, 
and there are pairs of 
points in ${\cal M}_{\rm ex}$
whose neighborhoods also overlap in every 
sector of~${\cal M}_{\rm reg}$. 
Finally, any neighborhood of ${\cal M}_0$ 
contains ${\cal M}_{\rm ex}$ 
and intersects all the sectors of~${\cal M}_{\rm reg}$. 

We therefore see that ${\cal M}$ is connected: 
each of the disconnected sectors 
of ${\cal M}_{\rm reg}$ is attached to 
${\cal M}_0$ and~${\cal M}_{\rm ex}$. 
It is clear from (\ref{Mgen-reps}) that 
the subset ${\cal M}_{\rm reg} \cup {\cal M}_0$ can be visualized as
four cones with a common tip, the tip 
consisting of 
the single point in ${\cal M}_0$ and being at 
$r\to0_+$ in each ${\cal M}_{\epsilon_1,\epsilon_2}$ \cite{monrovthie}. 
On the other hand, for fixed~$\epsilon_2$, the union of 
${\cal M}_{1,\epsilon_2}$, 
${\cal M}_{-1,\epsilon_2}$, and the $\theta=0$ circle of 
${\cal M}_{\rm ex}$ constitutes a smooth symplectic manifold with
topology $\BbbR\times S^1$: to see this, make in 
(\ref{Mgen-reps}) a gauge transformation that 
multiplies 
$\vec v$ and 
$\vec p$ by $\sqrt{r}$ and divides 
$\vec u$ and 
$\vec \pi$ by~$\sqrt{r}$, and allow $r$ to take all real values. 
The union of 
${\cal M}_{1,\epsilon_2}$, 
${\cal M}_{-1,\epsilon_2}$, and the $\theta=\pi/2$ circle of 
${\cal M}_{\rm ex}$ constitutes also a smooth symplectic manifold with
topology $\BbbR\times S^1$: 
to see this, make in 
(\ref{Mgen-reps}) the analogous gauge transformation with 
$1/\sqrt{r}$ instead of~$\sqrt{r}$. The union of 
${\cal M}_{1,\epsilon_2}$, 
${\cal M}_{-1,\epsilon_2}$, and both of these circles in 
${\cal M}_{\rm ex}$ is a smooth symplectic non-Hausdorff manifold,
with topology $\BbbR'\times
S^1$, where $\BbbR'$ is the real line with doubled origin. 
The structure of ${\cal M}$ near ${\cal M}_{\rm ex}$ is therefore 
reminiscent of, but more
involved than, the joining of the causal and noncausal 
sectors of Misner space 
\cite{haw-ell}, 
or the joining of the spacelike and timelike sectors in the solution
space to Witten's 2+1 gravity on $\BbbR\times T^2$
\cite{loumar-wittentorus,WitMatNi} or on 
$\BbbR\times$(Klein bottle) 
\cite{louko-wittenkb}. 

We now turn to the observables. Consider on 
$\Gamma$ the six functions \cite{monrovthie}
\begin{eqnarray}
O_{12} := u^{1}p^{2}-p^{1}u^{2}, &\hspace{2em}& O_{23} := 
u^{2}v^{1}-p^{2}\pi^{1}, \nonumber \\
O_{13} := u^{1}v^{1}-p^{1}\pi^{1}, &\hspace{2em}& O_{24} := 
u^{2}v^{2}-p^{2}\pi^{2}, \nonumber \\
O_{14} := u^{1}v^{2}-p^{1}\pi^{2}, &\hspace{2em}& O_{34} := 
\pi^{1}v^{2}-v^{1}\pi^{2}  
\ . 
\label{perennials}
\end{eqnarray}
The linear span of the $O_{ij}$ is closed under Poisson brackets,
and the Poisson bracket algebra is isomorphic to the Lie
algebra
$\otwotwo \simeq \sltwor \times \sltwor$. 
The basis (\ref{perennials}) is adapted to the $\otwotwo$ form of 
the algebra, while a basis adapted to the $\sltwor \times \sltwor$
form is 
\begin{eqnarray}
\tau_0^\eta 
&:=& 
\casehalf \left( O_{12} - \eta O_{34} \right)
\ \ ,
\nonumber
\\
\tau_1^\eta 
&:=& 
\casehalf \left( O_{13} - \eta O_{24} \right)
\ \ ,
\nonumber
\\
\tau_2^\eta 
&:=& 
\casehalf \left( O_{23} + \eta O_{14} \right)
\ \ ,
\label{tau-def}
\end{eqnarray}
where $\eta \in \{1,-1\}$: the Poisson brackets read
\begin{eqnarray}
\{ \tau_1^\eta  , \, \tau_2^{\eta'} \} 
& = & 
- \delta^{\eta, \eta'}\tau_0^{\eta}
\ \ ,
\nonumber
\\
\{ \tau_2^\eta  , \, \tau_0^{\eta'} \} 
& = & 
\delta^{\eta, \eta'}\tau_1^{\eta}
\ \ ,
\nonumber
\\
\{ \tau_0^\eta  , \, \tau_1^{\eta'} \} 
& = & 
\delta^{\eta, \eta'}\tau_2^{\eta}
\ \ . 
\end{eqnarray}
We record for future use that the $\tau^\eta_i$ satisfy for each $\eta$
the identity 
\begin{equation}
- \left( \tau_0^\eta \right)^2 
+ \left( \tau_1^\eta \right)^2 
+ \left( \tau_2^\eta \right)^2 
= 
H_1 H_2 + \case{1}{4} D^2 
\ \ .
\label{taus-identity}
\end{equation}

Now, $\tau^\eta_i$ Poisson commute with the constraints and are thus
by definition observables. We denote by $\Aclass$ the 
classical observable 
algebra generated by~$\{\tau^\eta_j\}$. 
The pullbacks of $\tau^\eta_i$ to ${\cal M}$
vanish on ${\cal M}_0$ and~${\cal M}_{\rm ex}$, 
while on ${\cal M}_{\rm reg}$ we have 
\begin{mathletters}
\label{taus-on-calM}
\begin{eqnarray}
&&
\tau^\eta_0
= 
\casehalf \epsilon_1 
(1 + \eta\epsilon_2)
\, 
r 
\ \ ,
\label{tauzero-on-calM}
\\
&&
\tau^\eta_1
= 
\casehalf 
(1 + \eta\epsilon_2)
\, 
r \cos\varphi
\ \ ,
\\
&&
\tau^\eta_2
= 
- \casehalf \epsilon_1 
(1 + \eta\epsilon_2)
\, 
r \sin\varphi
\ \ .
\end{eqnarray}
\end{mathletters}
$\Aclass$ therefore separates~${\cal M}_{\rm reg}$. More precisely,
for given~$\eta$, the $\sltwor$ subalgebra generated by
$\{\tau^\eta_i\}$ vanishes on 
${\cal M}_{1,-\eta}$ and ${\cal M}_{-1,-\eta}$
but separates 
${\cal M}_{1,\eta} \cup {\cal M}_{-1,\eta}$, 
and on
${\cal M}_{\epsilon_1,\eta}$ 
$\tau^\eta_0$ has 
the definite sign~$\epsilon_1$.

We note in passing that 
$\tau^\eta_i$ are real
analytic functions on~$\Gamma$. 
For given $\epsilon_1$ and~$\epsilon'_1$, 
(\ref{taus-on-calM}) therefore shows that 
${\cal M}_{\epsilon_1,1}$ and ${\cal M}_{\epsilon_1',-1}$ cannot
both belong to a connected real analytic manifold whose analytic 
structure would be induced from that of~$\Gamma$.

By construction, exponentiating the Poisson bracket action of 
$\Aclass$ on $\Gamma$ yields on $\Gamma$ 
the action of a connected group ${\cal G}$ that is locally 
$\SLtwor\times\SLtwor$, 
and this ${\cal G}$-action commutes with the gauge group
action~(\ref{class-gaugeaction}). 
Considering $\Gamma$ in a
polarization in which $(\vec u, \vec \pi)$ are the `coordinates' and 
$(\vec p, -\vec v)$ are the `momenta', it is immediate from
(\ref{perennials}) that this ${\cal G}$-action reads 
\begin{mathletters}
\label{Otwotwo-caction}
\begin{eqnarray}
\pmatrix{\vec u \cr \vec \pi \cr}
&&
\mapsto 
A 
\pmatrix{\vec u \cr \vec \pi \cr}
\ \ ,
\label{Otwotwo-upi-caction}
\\
\pmatrix{\vec p \cr -\vec v \cr}
&&
\mapsto 
{(A^{-1})}^T
\pmatrix{\vec p \cr -\vec v \cr}
\ \ ,
\end{eqnarray}
\end{mathletters}
where $A$ is a $4\times4$ matrix in the defining representation of
$\Otwotwo$, and in the connected component~$\Octwotwo$. 
Hence ${\cal G} = \Octwotwo \simeq 
[\SLtwor\times\SLtwor]/\BbbZ_2$. We use 
(\ref{Otwotwo-caction}) to extend the 
${\cal G}$-action to the action of 
${\cal G}_+ := \Otwotwo$: the 
${\cal G}_+$-action is generated by the ${\cal G}$-action and the 
four maps 
$P_{\epsilon_1, \epsilon_2}$, 
where $\epsilon_i \in \{1,-1 \}$ and 
\begin{equation}
P_{\epsilon_1, \epsilon_2}: 
(u^1, u^2, v^1, v^2, 
p^1, p^2, \pi^1, \pi^2) 
\mapsto 
(u^1, \epsilon_1 u^2, v^1, \epsilon_1 \epsilon_2 v^2, 
p^1, \epsilon_1 p^2, \pi^1, \epsilon_1 \epsilon_2 \pi^2) 
\ \ . 
\label{Pclass-defs}
\end{equation}
It is clear that also the ${\cal
  G}_+$-action on $\Gamma$ commutes with the gauge group
action~(\ref{class-gaugeaction}). 

The induced ${\cal G}$-action on ${\cal M}$ is trivial on 
${\cal M}_0$, maps ${\cal M}_{\rm ex}$ transitively to itself, 
and maps each 
${\cal M}_{\epsilon_1, \epsilon_2}$ transitively to itself. 
The induced ${\cal G}_+$-action on ${\cal M}$ is trivial
on~${\cal M}_0$, maps ${\cal M}_{\rm ex}$ transitively to itself, 
and maps 
${\cal M}_{\rm reg}$ transitively to itself, permuting the 
the four sectors of ${\cal M}_{\rm
  reg}$ by a $\BbbZ_2\times\BbbZ_2$ permutation subgroup according to 
\begin{equation}
P_{\epsilon'_1, \epsilon'_2}: 
{\cal M}_{\epsilon_1, \epsilon_2}
\to {\cal M}_{\epsilon'_1 \epsilon_1, \epsilon'_2 \epsilon_2}
\ \ .
\label{Pclass-permutation}
\end{equation}

\section{Algebraic Quantization}
\label{sec:oldAQ}

In this section we quantize the system in the Algebraic Quantization
framework of \cite{AAbook}.  In this framework one first solves the
quantum constraint equations on a linear space, without an inner
product, and then seeks a Hermitian inner product such that the
adjoint relations in the chosen quantum observable algebra
reflect the reality relations in the corresponding classical
observable algebra; we refer to \cite{AAbook,ashtate} for
overviews and more detail.  Our analysis closely follows that in 
\cite{monrovthie}, the main difference being that
we consider two possible choices for the classical observable algebra,
arising respectively from the groups ${\cal G}$ and ${\cal
  G}_+$ introduced in section~\ref{sec:classical}. 
The connection to Isham's group
theoretic quantization \cite{isham-les} is made at the end of the
section. 

We work in a ``coordinate representation'', starting with the
linear space of smooth functions $\Psi({\vec u},{\vec v})$ on $\BbbR^4$. 
We shall frequently use the polar coordinates defined by 
$u^1 + i u^2 = u e^{i\alpha}$, $v^1 + i v^2 = v
e^{i\beta}$, where $u\ge0$, $v\ge0$. 
Note that no inner product is introduced at this stage. 

To begin, we promote the classical constraints (\ref{constraints})
into quantum constraint operators. The momentum operators are
\begin{equation}
{\widehat {\vec p}} := -i \vec \nabla_{u}, \ \ \ 
{\widehat {\vec \pi}} := -i \vec \nabla_{v} \ \ ,
\label{oper}
\label{naive-momentum}
\end{equation}
and we order the quantum constraints as 
\begin{mathletters}
\label{q-constraints}
\begin{eqnarray}
{\widehat H}_1 & := & 
- \casehalf \left( 
\vec\nabla_{u}^{2}
+ \vec v^2 \right)
\ \ ,
\\
{\widehat H}_2 & := & 
- \casehalf \left( 
\vec\nabla_{v}^{2}
+ \vec u^2 \right)
\ \ ,
\\
{\widehat D} & := & 
-i 
\left( 
{\vec u} \cdot \vec\nabla_{u}
- {\vec v} \cdot \vec\nabla_{v} \right)
\ \ ,
\end{eqnarray}
\end{mathletters}
where $\vec\nabla_{u}^{2} := 
{\partial^{2}\over\partial{(u^{1})^{2}}}\ 
+{\partial^{2}\over\partial{(u^{2})^{2}}}$, 
and similarly for $\vec\nabla_{v}^{2}$. The commutator algebra of the
quantum constraints then closes as 
\begin{mathletters}
\label{q-constraints-alg}
\begin{eqnarray}
\left[ {\widehat H}_1 \, , \, {\widehat H_2} \right]
& = & 
i {\widehat D}
\ \ ,
\\
\left[ {\widehat D} \, , \, {\widehat H}_1 \right] 
& = & 
2 i {\widehat H}_1
\ \ ,
\\
\left[ {\widehat D} \, , \, {\widehat H}_2 \right] 
& = & 
- 2 i {\widehat H}_2
\ \ .
\end{eqnarray}
\end{mathletters}

Next, we define a set of quantum observables ${\widehat O}_{ij}$ by
substituting the momentum operators (\ref{oper}) into 
the expressions (\ref{perennials}) of the classical
observables~$O_{ij}$. 
As the resulting expressions 
contain no products of
noncommuting operators, no issue of ordering arises. 
The operators 
${\widehat O}_{ij}$ commute with the
constraints~(\ref{q-constraints}), and 
their commutator algebra closes. 
As $O_{ij}$ are real, we introduce on this 
algebra a star-operation by 
${\widehat O}_{ij}^{\star} =
{\widehat O}_{ij}$ 
and extending to the full algebra by
antilinearity. We denote this star-algebra of physical observables 
by~$\Aphystar$. 

We define in $\Aphystar$ 
the operators ${\widehat\tau}_i^\eta$ by the
hatted counterparts
of~(\ref{tau-def}), and we write 
\begin{equation}
{\widehat\tau}_\pm^\eta
:= 
{\widehat\tau}_1^\eta \pm i {\widehat\tau}_2^\eta
\ \ .
\end{equation}
The operators ${\widehat\tau}_0^\eta$ and ${\widehat\tau}_\pm^\eta$
generate~$\Aphystar$. The commutators are
\begin{mathletters}
\label{taus-qcomms}
\begin{eqnarray}
{[ {\widehat\tau}_0^\eta  , \, {\widehat\tau}_\pm^{\eta'} ] }
& = & 
\pm \delta^{\eta, \eta'} \, {\widehat\tau}_\pm^\eta
\ \ ,
\\
{[  {\widehat\tau}_+^\eta  , \, {\widehat\tau}_-^{\eta'} ] }
& = & 
- 2 \delta^{\eta, \eta'} \, {\widehat\tau}_0^\eta
\ \ , 
\end{eqnarray}
\end{mathletters}
and the 
star-operation reads 
\begin{mathletters}
\label{tau-starrels}
\begin{eqnarray}
{({\widehat\tau}_0^\eta)}^{\star} 
&=& 
{\widehat\tau}_0^\eta
\ \ ,
\\
{({\widehat\tau}_\pm^\eta)}^{\star} 
&=& 
{\widehat\tau}_\mp^\eta
\ \ .
\end{eqnarray}
\end{mathletters}
The explicit expressions of the operators in the polar coordinates
are
\begin{mathletters}
\label{polartaus}
\begin{eqnarray}
{\widehat\tau}_0^\eta
&=&
- \casehalf i 
\left( \partial_\alpha + \eta \partial_\beta \right)
\ \ ,
\\
{\widehat\tau}_\pm^\eta
&=&
\casehalf e^{\pm i (\alpha + \eta \beta)}
\left\{
uv + 
\left[ \partial_u \pm (i/u) \partial_\alpha \right]
\left[ \partial_v \pm \eta(i/v) \partial_\beta \right] 
\right\}
\ \ .
\end{eqnarray}
\end{mathletters}

We now solve the quantum constraints by separation of variables. 
As shown in \cite{monrovthie}, solutions that are smooth functions 
of $(\vec u, \vec v)$ and separable in their angle dependence
are multiples of the functions 
\begin{equation}
\Psi_{m,\epsilon} := 
e^{i m(\alpha + \epsilon \beta)} \ 
J_m (uv)
\ \ ,
\label{q-solutions}
\end{equation}
where $m\in\BbbZ$, $\epsilon \in \{1,-1\}$, and $J_m$
is the Bessel function of the first kind 
\cite{Grad-Rhyz}. 
The functions 
$\Psi_{m,\epsilon}$ are linearly independent, with the exception that 
$\Psi_{0,+} = \Psi_{0,-}$. We denote the linear span of the
$\Psi_{m,\epsilon}$ by~${\tilde V}$. As 
\begin{mathletters}
\label{astarrepV}
\begin{eqnarray}
&&{\widehat\tau}_0^\eta \Psi_{m,\epsilon} 
= 
\delta^{\eta,\epsilon} m 
\Psi_{m,\epsilon} 
\ \ ,
\label{astarrepV-nought}
\\
&&{\widehat\tau}_\pm^\eta \Psi_{m,\epsilon} 
= \delta^{\eta,\epsilon} m 
\Psi_{m\pm1,\epsilon} 
\ \ ,
\label{astarrepV-pm}
\end{eqnarray}
\end{mathletters}
${\tilde V}$
carries a representation of~$\Aphystar$. 

One could now find the subspaces of ${\tilde V}$ on which the
representation of $\Aphystar$ is (algebraically) 
irreducible, 
and look on each for an
inner product in which the 
star-operation (\ref{tau-starrels}) becomes
the adjoint operation, 
\begin{mathletters}
\label{tau-hermrels}
\begin{eqnarray}
{({\widehat\tau}_0^\eta)}^{\dagger} 
&=& 
{\widehat\tau}_0^\eta
\ \ ,
\label{taunought-hermrels}
\\
{({\widehat\tau}_\pm^\eta)}^{\dagger} 
&=& 
{\widehat\tau}_\mp^\eta
\ \ .
\label{taupm-hermrels}
\end{eqnarray}
\end{mathletters}
However, the only subspace on which such an inner product exists is
the one-dimensional subspace generated by~$\Psi_{0,+}$, and the
resulting theory is physically uninteresting, as every operator in 
$\Aphystar$ then 
annihilates the whole Hilbert space. There are four other subspaces
carrying an irreducible 
representation of~$\Aphystar$, 
but each of these subspaces contains~$\Psi_{0,+}$, 
and the adjoint relations (\ref{tau-hermrels}) 
imply that $\Psi_{0,+}$ have a 
vanishing norm [cf.~(\ref{recurrence}) 
and (\ref{norm-on-Vs}) below]. 

The way to remedy the situation is to note that the troublesome vector 
$\Psi_{0,+}$ is annihilated by every operator
in~$\Aphystar$, and this vector can therefore be dropped at the
outset. 
Let thus $V$ be the linear span of
$\left\{\Psi_{m,\epsilon} \mid m\ne0\right\}$. 
$V$~carries a representation of~$\Aphystar$, which reads as 
in (\ref{astarrepV}) except that whenever $\Psi_{0,\epsilon}$ 
would occur on the right-hand side, it is replaced by the zero
vector. $V$ decomposes into the direct sum $V = 
\bigoplus V_{\epsilon_1,\epsilon_2}$, 
where $\epsilon_i \in \{1,-1\}$ and 
\begin{equation}
V_{\epsilon_1,\epsilon_2} :=
{\mbox{span}} \left\{ \Psi_{m,\epsilon_2} \mid \epsilon_1 m > 0 \right\}
\ \ . 
\label{Vs}
\end{equation}
Each $V_{\epsilon_1,\epsilon_2}$ carries an irreducible representation 
of~$\Aphystar$, and we therefore seek an inner product
${(\,\cdot\,,\,\cdot\,)}_{\epsilon_1,\epsilon_2}$
individually on each. 
Equations (\ref{astarrepV-nought}) and (\ref{taunought-hermrels})
imply that the $\Psi_{m,\epsilon}$ are orthogonal. Equations
(\ref{astarrepV-pm}) and (\ref{taupm-hermrels}) yield the recurrence
relation
\begin{eqnarray}
(m\pm1)^2 \left( \Psi_{m} \, , \Psi_{m} \right)
&=&
\left( 
{\widehat\tau}_\mp \Psi_{m\pm1} \, , 
{\widehat\tau}_\mp \Psi_{m\pm1} 
\right)
\nonumber
\\
&=&
\left( 
\Psi_{m\pm1} \, , 
{\widehat\tau}_\pm
{\widehat\tau}_\mp
\Psi_{m\pm1} 
\right)
\nonumber
\\
&=&
m(m\pm1)
\left( 
\Psi_{m\pm1} \, , 
\Psi_{m\pm1} 
\right)
\ \ ,
\label{recurrence}
\end{eqnarray}
where we have suppressed the index $\epsilon$ on the vectors, the
index $\eta=\epsilon$ on ${\widehat\tau}_\pm$, and the index on the 
inner product. It follows, still suppressing the indices, that 
\begin{equation}
{\left( \Psi_{m} \, , \Psi_{m'} \right)}
= a |m| \delta_{m, m'}
\ \ ,
\label{norm-on-Vs}
\end{equation}
where $a$ is a positive constant, independent for each
$V_{\epsilon_1,\epsilon_2}$. 

It is clear that (\ref{norm-on-Vs}) defines on each
$V_{\epsilon_1,\epsilon_2}$ an inner product satisfying the adjoint
relations~(\ref{tau-hermrels}). 
Completion yields the four 
Hilbert spaces~${\cal H}_{\epsilon_1,\epsilon_2}$, 
and it follows from the asymptotic large
order expansion of $J_m$ \cite{abra-stegun} that every vector in these 
Hilbert spaces is represented by a function on the original
configuration space $\BbbR^4 = \left\{(\vec u, \vec v)\right\}$. 
Each ${\cal H}_{\epsilon_1,\epsilon_2}$ 
carries 
a representation of $\Aphystar$ by densely-defined operators. 
For given~$\eta$, the 
$\sltwor$ subalgebra generated by 
$\{{\widehat\tau}^\eta_i\}$ is represented nontrivially 
on~${\cal H}_{\epsilon_1,\eta}$: the representation belongs to the
discrete series \cite{howe-tan,bargmann,lang,knapp} 
and, in the notation of \cite{bargmann}, 
is known as~$D^{\epsilon_1}_1$. 

In each of these representations of 
$\Aphystar$ on~${\cal H}_{\epsilon_1,\epsilon_2}$, 
the Casimir operators of both the
trivial and nontrivial $\sltwor$
subalgebra take the value zero: 
\begin{equation}
\left[
- \left( {\widehat\tau}_0^\eta \right)^2 
+ \left( {\widehat\tau}_1^\eta \right)^2 
+ \left( {\widehat\tau}_2^\eta \right)^2 
\right]
{\cal H}_{\epsilon_1,\epsilon_2}
= 0 
\ \ .
\label{taus-casimir}
\end{equation}
In this sense, the quantum theory has preserved the identities 
(\ref{taus-identity}) satisfied by the 
classical observables. 

It is easy to extend the above analysis to the larger observable
algebra~$\Aphyplusstar$, generated by $\Aphystar$ 
and the set 
$\left\{{\widehat P}_{\epsilon_1, \epsilon_2}\right\}$, 
where $\epsilon_i \in \{1,-1 \}$ and 
\begin{equation}
\left({\widehat P}_{\epsilon_1, \epsilon_2} \Psi \right)
(u^1, u^2, v^1, v^2 ) 
:= \Psi (u^1, \epsilon_1 u^2, v^1, \epsilon_1 \epsilon_2 v^2 ) 
\ \ . 
\label{P-defs}
\end{equation}
Note that ${\widehat P}_{\epsilon_1, \epsilon_2}$ is the operator
analogue of the map $P_{\epsilon_1, \epsilon_2}$ (\ref{Pclass-defs})
on~$\Gamma$. 
The star-operation
is extended to 
$\Aphyplusstar$ by 
${\widehat P}_{\epsilon_1, \epsilon_2}^{\star} 
= {\widehat P}_{\epsilon_1, \epsilon_2}$. 
As 
\begin{equation}
{\widehat P}_{\epsilon_1, \epsilon_2}
\Psi_{m,\epsilon} 
= 
\Psi_{\epsilon_1 m, \epsilon_2 \epsilon} 
\ \ ,
\label{P-repV}
\end{equation}
the new operators permute the subspaces $V_{\epsilon_1, \epsilon_2}$
by an $\BbbZ_2\times\BbbZ_2$ permutation subgroup 
according to 
\begin{equation}
{\widehat P}_{\epsilon'_1, \epsilon'_2}
V_{\epsilon_1, \epsilon_2}
= 
V_{\epsilon'_1 \epsilon_1, \epsilon'_2 \epsilon_2}
\ \ ,
\end{equation}
and the representation of $\Aphyplusstar$ on $V$
is irreducible.
Proceeding as above, we arrive at the 
Hilbert space ${\cal H}_+
:= \bigoplus {\cal H}_{\epsilon_1, \epsilon_2}$, 
where the subspaces ${\cal H}_{\epsilon_1, \epsilon_2}$ are
orthogonal and 
the inner product on each is given by~(\ref{norm-on-Vs}), 
but now with the same $a$ for all~${\cal H}_{\epsilon_1,
  \epsilon_2}$. 

The quantum theories that we have obtained have a natural
interpretation as quantizations of different subsets of the classical
reduced phase space~${\cal M}$.  
For given $\epsilon_1$ and~$\epsilon_2$, 
the representation of $\Aphystar$ on
${\cal H}_{\epsilon_1,\epsilon_2}$ is the quantum analogue of the 
pullback of the
classical algebra $\Aclass$ to ${\cal M}_{\epsilon_1,\epsilon_2}$, in
that in each case the $\eta=-\epsilon_2$ $\sltwor$ subalgebra is
trivial, and in the nontrivial $\sltwor$ subalgebra
${\widehat\tau}_0^{\epsilon_2}$ and and ${\tau}_0^{\epsilon_2}$ have
the same definite sign.  The Hilbert space ${\cal H}_{\epsilon_1,
  \epsilon_2}$ with the observable algebra $\Aphystar$ can therefore
be thought of as a quantization of the sector ${\cal M}_{\epsilon_1,
  \epsilon_2}$.  Similarly, the Hilbert space ${\cal H}_+$ with the
observable algebra $\Aphyplusstar$ can be thought of as a quantization
of all the four sectors of~${\cal M}_{\rm reg}$.

One can also obtain our quantum theories via the group theoretic
quantization of Isham \cite{isham-les}. As noted in
section~\ref{sec:classical}, the ${\cal G}$-action 
(\ref{Otwotwo-caction}) on $\Gamma$ induces on each 
${\cal M}_{\epsilon_1,\epsilon_2}$ a transitive ${\cal G}$-action,
and also the transitive action of a 
subgroup $\SLtwor\subset {\cal G}$:
this $\SLtwor$-action is obtained by
exponentiating 
the Poisson bracket action of the algebra~(\ref{taus-on-calM}). 
For group 
theoretic quantization on 
a given sector~${\cal M}_{\epsilon_1,\epsilon_2}$, we can therefore 
adopt this $\SLtwor$ as the canonical group. 
In order 
to preserve the classical 
identities 
(\ref{taus-identity}) in the quantum theory, we consider the 
irreducible unitary 
representations of $\SLtwor$ in which the Casimir operator 
vanishes. 
The only such representations are the trivial representation and the
discrete series representations 
$D^{\pm}_1$ \cite{howe-tan,bargmann,lang,knapp}. 
${\widehat\tau}_0^{\epsilon_2}$~vanishes in the trivial
representation, whereas in each $D^{\pm}_1$ it as a definite sign, and 
it is in $D^{\epsilon_1}_1$ that this sign agrees with the sign of the 
classical function ${\tau}_0^{\epsilon_2}$
(\ref{tauzero-on-calM}) on~${\cal M}_{\epsilon_1,\epsilon_2}$. 
Thus, requiring the signs of ${\widehat\tau}_0^{\epsilon_2}$ and
${\tau}_0^{\epsilon_2}$ to agree picks the representation 
$D^{\epsilon_1}_1$: we arrive at
the Hilbert space~${\cal
  H}_{\epsilon_1,\epsilon_2}$,
and the observable algebra is the $\sltwor$ subalgebra of 
$\Aphystar$ with $\eta=\epsilon_2$. 
A~similar argument can be made 
for group theoretic quantization on ${\cal M}_{\rm reg}$ with the
canonical group ${\cal G}_+ \simeq \Otwotwo \simeq \Octwotwo \times_s
(\BbbZ_2)^2$, arriving at~${\cal H}_+$ 
with the observable algebra~$\Aphyplusstar$. As neither ${\cal M}_{\rm
  reg}$ nor ${\cal G}_+$ is connected, it is perhaps not clear how
unique the implementation of the group theoretic quantization in
this case is, but ${\cal H}_+$ clearly 
does carry an
irreducible unitary representation of~${\cal G}_+$. 
Further possibilities of implementing group theoretic quantization on 
${\cal M}_{\rm
  reg}$ and its four sectors are discussed in
\cite{bojo-etal,bojo-strobl}. 

We end the section with two remarks: 

1) 
One might have tried to include in 
the vector space of solutions to the
constraints functions that are not smooth at $uv=0$. In
this case one can replace $J_m$ in (\ref{q-solutions}) by any linear
combination of $J_m$ and~$N_m$, with $m$-independent coefficients, and
the abstract construction of the Hilbert spaces goes through 
as above.  However, when $N_m$ is present, it is seen from the 
large order expansion of $N_m$ \cite{abra-stegun} that the completion
introduces in the Hilbert spaces vectors that cannot be represented by
functions on the original configuration space.

2) 
One might have tried to include in 
the vector space of solutions to the constraints 
vectors that are not single-valued functions on the
configuration space, thus allowing $m$ in~(\ref{q-solutions}), or in
the analogue of (\ref{q-solutions}) with a linear combination of 
$J_m$ and~$N_m$, to be noninteger. The representation of $\Aphystar$
on this larger vector space takes again the form 
(\ref{astarrepV}), and breaks thus into irreducible representations
classified by $\epsilon$ and the fractional part of~$m$. However, in
this case no inner product satisfying the adjoint relations
(\ref{tau-hermrels}) exists.

\section{Formalism of Refined Algebraic Quantization with 
group averaging}
\label{sec:RAQoutline}

In this section we give a brief outline of Refined Algebraic
Quantization (RAQ) with group averaging. The main purposes of the
section are to fix the notation and to fix the particular version
of RAQ: we follow the formulation of 
Giulini
and Marolf \cite{giumar-unique}. 
We specialize throughout to the case where the
gauge group is a connected unimodular Lie group.

\subsection{Refined Algebraic Quantization}
\label{subsec:RAQgeneral}

RAQ begins by implementing the quantum constraints as self-adjoint
operators on an auxiliary Hilbert space~$\Haux$. We
assume that the commutator algebra of the constraints closes as a Lie
algebra, so that the algebra exponentiates into a unitary
representation $U(g)$ of a corresponding connected Lie group 
$G$ on~$\Haux$. We refer to
$G$ as the gauge group, and we assume that it is unimodular (that is,
that the structure constants of the Lie algebra are traceless). 

Next, RAQ solves the constraints in an 
enlargement of~$\Haux$. To this end, one
introduces a space of test states, 
a dense linear subspace $\Phi\subset\Haux$ such that the
operators $U(g)$ map $\Phi$ to itself. The desired enlargement is
the algebraic dual of~$\Phi$, denoted by $\Phi^*$ and topologized 
by the topology of pointwise convergence. For $f\in\Phi^*$ and
$\phi\in\Phi$, we denote the dual action of $f$ on $\phi$
by~$f[\phi]$. 
$\Phi^*$~carries a
representation $U^*(g)$ of 
$G$ defined by the dual action: for 
$f\in\Phi^*$, 
$\biglb(U^*(g)f\bigrb)[\phi] = f[U(g^{-1})\phi]$ for all $\phi \in
\Phi$. Solutions to the quantum constraints are then by definition
the elements 
$f\in \Phi^*$ for which $U^*(g)f = f$ for all $g\in G$. 

The RAQ algebra of observables is completely
determined by the structure specified above. An operator ${\cal O}$ on
$\Haux$ is called gauge invariant if the domains of 
${\cal O}$ and ${\cal O}^\dag$ 
include~$\Phi$, ${\cal O}$ and ${\cal O}^\dag$ 
map $\Phi$ to itself, and ${\cal O}$ 
commutes with the $G$-action on~$\Phi$:
${\cal O} U(g) \phi = U(g) {\cal O} \phi$ for all $g\in G$,
$\phi\in\Phi$. 
Note that if ${\cal O}$ is gauge invariant, so is~${\cal O}^\dag$. 
The observable algebra 
$\Aobs$ is by definition the algebra of
gauge invariant operators. 
$\Aobs$~has on $\Phi^*$ an antilinear representation
defined by the dual action \cite{giumar-general}: 
for $f\in\Phi^*$, $({\cal O}f)[\phi] = f [{\cal O}^\dag \phi]$
for all $\phi \in \Phi$. 
Note that $\Aobs$ 
does not need to be constructed or presented in any explicit
sense. 

The last ingredient in RAQ is a rigging map, which is by definition an
antilinear map $\eta$ from $\Phi$ to $\Phi^*$ satisfying 
four postulates:

(i)~The image of $\eta$ solves the constraints: Each vector in the
image of $\eta$ is invariant under the $G$-action on~$\Phi^*$.

(ii)~$\eta$ is real: 
$\eta(\phi_1)[\phi_2] = \overline{\eta(\phi_2)[\phi_1]}$ 
for all $\phi_1, \phi_2 \in \Phi$. 

(iii)~$\eta$ is positive: 
$\eta(\phi)[\phi] \ge 0$
for all $\phi\in \Phi$. 

(iv)~$\eta$ intertwines with the 
representations of the observable algebra on $\Phi$ and~$\Phi^*$: 
${\cal O} ( \eta \phi ) = \eta ({\cal O} \phi)$ for all 
${\cal O}\in\Aobs$ and all $\phi\in\Phi$. 

The input required in RAQ is now complete. As the final step, RAQ 
introduces on 
the image of $\eta$ a Hermitian inner
product by 
\begin{equation}
\biglb(\eta(\phi_1), \eta(\phi_2) \bigrb)_{\rm phys} 
:= \eta(\phi_2)[\phi_1]
\ \ , 
\label{phys-ip}
\end{equation}
and completes the image of $\eta$ in this inner product into a Hilbert 
space~$\Hphys$, which is by definition 
the physical Hilbert space of the theory. 
$\Hphys$~carries an antilinear representation of~$\Aobs$, 
and the
adjoint map in this representation (with respect to the inner product
on $\Hphys$) is by construction 
that induced by the adjoint map on~$\Haux$. The
representation of $\Aobs$ on $\Hphys$ is known to be nontrivial
provided certain technical conditions hold \cite{giumar-general}.

\subsection{Group averaging}
\label{subsec:RAQaveraging}

The group averaging proposal in RAQ addresses the last ingredient
above, the choice of the rigging map. The proposal 
seeks the rigging map as a suitable interpretation of the formal 
expression 
\begin{equation}
\eta (|\phi\rangle)
:= 
\int_G dg \, 
\langle \phi| 
U(g)
\ \ ,
\label{ga1}
\end{equation}
where we have invoked the Dirac notation for the vector
$|\phi\rangle\in\Phi$ and for its Hilbert dual vector~$\langle\phi|$. 
The measure $dg$ is the Haar measure on $G$ 
(which is both left and right invariant 
by the unimodularity of~$G$). 

Consider now the formula
\begin{equation}
{(\phi_2 , \phi_1 )}_{\rm ga} := 
\int_G d g \, \biglb(\phi_2 , U(g) \phi_1 \bigrb)_{\rm aux}
\ \ , 
\label{GAM}
\end{equation}
and suppose that the integral on the right-hand side 
converges in absolute value for all 
$\phi_1$ and $\phi_2$ in~$\Phi$. Formula (\ref{GAM}) defines then on
$\Phi$ the 
sesquilinear form 
${(\,\cdot\,,\,\cdot\,)}_{\rm ga}$, and we
interpret the group averaging proposal 
(\ref{ga1}) as 
\begin{equation}
\eta(\phi_1)[\phi_2] 
:= 
{(\phi_1 , \phi_2 )}_{\rm ga}
\ \ . 
\label{GAM-eta}
\end{equation}
The resulting map $\eta$ clearly satisfies 
postulates (i), (ii), and~(iv): (i)~follows from the
invariance of the Haar measure, and (ii) from the
fact that $dg = d(g^{-1})$. 
If $\eta$ further satisfies~(iii), 
and if $\eta$ is not identically zero, 
the group
averaging proposal has then produced a rigging map. 


Considerable control over the space of possible rigging maps is
provided by the uniqueness theorem of Giulini and Marolf
\cite{giumar-unique}.  To state the theorem, we note
\cite{giumar-unique} that 
if $h$ is an $L^1$ function on~$G$, the expression
${\hat{h}} := \int_G dg \, h(g) U(g)$ defines a bounded operator
on~$\Haux$, and the set of all such operators forms an
algebra~$\Agrouphat$. Suppose now that
$\Phi$ is invariant under~$\Agrouphat$, the integral in (\ref{GAM})
converges in absolute value for all $\phi_1$ and $\phi_2$ in~$\Phi$,
and the sesquilinear form 
${(\,\cdot\,,\,\cdot\,)}_{\rm ga}$ on $\Phi$ is not
identically zero.
Then, if a rigging map exists, it is unique up to
an overall multiple, and given by~(\ref{GAM-eta}) 
\cite{giumar-unique}.

\section{Refined Algebraic quantization 
of the $\SLtwor$ system}
\label{sec:RAQus}

In this section we apply the RAQ formalism of in section
\ref{sec:RAQoutline} to our system. 
To maintain a contact to the Algebraic Quantization
of section~\ref{sec:oldAQ}, 
we shall proceed so that the RAQ observable algebra $\Aobs$ 
will turn out to contain the Algebraic Quantization 
observable algebra~$\Aphyplusstar$.

\subsection{Auxiliary Hilbert space and 
the gauge group} 

We take the auxiliary Hilbert space $\Haux$ to be 
$L^2(\BbbR^4)$ 
of wave functions 
$\Psi({\vec u},{\vec v})$ in the inner product 
\begin{equation}
(\Psi_1,\Psi_2)_{\rm aux} := 
\int d^2\vec u \, d^2\vec v \, {\overline \Psi_1}\Psi_2
\ \ . 
\end{equation}
We take the constraint operators to be given by~(\ref{q-constraints}).

The constraints are essentially self-adjoint on~$\Haux$, and
exponentiating $-i$ times their algebra yields on $\Haux$ a unitary
representation $U$ of the universal covering group of $\SLtwor$.  The
group elements that appear in the Iwasawa decomposition
(\ref{iwa-abs}) are represented by
\begin{mathletters}
\label{Us-iwa}
\begin{eqnarray}
U \biglb( \exp (\beta e^- ) \bigrb) 
&=& 
\exp (-i \mu {\widehat H}_2 )
\ \ ,
\\
U \biglb( \exp(\lambda h) \bigrb) 
&=& 
\exp \biglb( -i \lambda {\widehat D} \bigrb) 
\ \ ,
\\
U \biglb( \exp[\theta(e^+ - e^-)] \bigrb) 
&=& 
\exp \biglb(-i \theta ( {\widehat H}_1 - {\widehat H}_2) \bigrb)
\ \ . 
\end{eqnarray}
\end{mathletters} 
$\exp (-i \mu {\widehat H}_2 )$ and 
$ \exp \biglb( -i \lambda {\widehat D} \bigrb)$
act on the wave functions
$\Psi({\vec u},{\vec v})$ respectively as 
\begin{mathletters}
\label{nullanddila-action}
\begin{eqnarray}
{[ \exp (-i \mu {\widehat H}_2 ) \, \Psi ]({\vec u},{\vec v})}
&=& 
\int
\frac{d^2 \vec v'} {2\pi i \mu} 
\exp \left\{
\frac{i}{2}
\left[
\frac{{(\vec v - \vec v')}^2}{\mu} 
+ \mu  \vec u^2 \right]
\right\}
\Psi({\vec u},{\vec v'})
\ \ \ 
\hbox{(for $\mu\neq0$)}
\ \ ,
\label{null-action}
\\
{[ \exp (-i \lambda {\widehat D}) \, \Psi ]({\vec u},{\vec v})}
&=& 
\Psi (e^{-\lambda}{\vec u},e^{\lambda}{\vec v})
\ \ . 
\label{dilation-action}
\end{eqnarray}
\end{mathletters}
Regarding 
$\exp \biglb(-i \theta ( {\widehat H}_1 - {\widehat H}_2) \bigrb)$, 
it suffices to observe that 
\begin{equation}
{\widehat H}_1 - {\widehat H}_2 
= {\widehat H}^{\rm sho}_{\vec u} - 
{\widehat H}^{\rm sho}_{\vec v}
\ \ ,
\label{comgen-rewrite}
\end{equation}
where 
${\widehat H}^{\rm sho}_{\vec u}$ and 
${\widehat H}^{\rm sho}_{\vec v}$ are the two-dimensional harmonic
oscillator Hamiltonians in respectively ${\vec u}$ and~${\vec v}$, 
\begin{mathletters}
\label{harmoscuv-hams}
\begin{eqnarray}
{\widehat H}^{\rm sho}_{\vec u} 
&:=& 
\casehalf \left( 
- \vec\nabla_{u}^{2}
+ \vec u^2 \right)
\ \ ,
\\
{\widehat H}^{\rm sho}_{\vec v} 
&:=& 
\casehalf \left( 
- \vec\nabla_{v}^{2}
+ \vec v^2 \right)
\ \ .
\end{eqnarray}
\end{mathletters}
It follows that 
$\exp \biglb(-i \theta 
( {\widehat H}_1 - {\widehat H}_2) \bigrb)$ is
periodic in $\theta$ with period~$2\pi$. As discussed in
appendix~\ref{app:iwasawa}, this 
shows that $U$ is a representation of $\SLtwor$ [and not just a
representation of the universal covering group of~$\SLtwor$]. 
In the terminology
of~RAQ, the gauge group $G$ is thus $\SLtwor$. 

The Algebraic Quantization observable algebra
$\Aphyplusstar$ is represented on 
$\Haux$ by densely-defined operators, and the
star-operation of $\Aphyplusstar$ is the adjoint map of~$\Haux$.
$\Aphyplusstar$~clearly commutes both with the constraint operators
(\ref{q-constraints}) and with $U$ on the respective
common domains. 
$\Aphyplusstar$~exponentiates into an $\Otwotwo$ action on~$\Haux$: 
representing the states as functions of 
$(\vec u, \vec \pi)$ via the Fourier-transform in~$\vec v$, 
$\Otwotwo$ acts on the
arguments of the functions
by~(\ref{Otwotwo-upi-caction}). It is clear that this 
$\Otwotwo$ action commutes with~$U$. 

$U$~is isomorphic to the 
oscillator representation of $\SLtwor$ on $L^2(\BbbR^{2,2})$, and our 
$\Otwotwo$ action on $\Haux$ is isomorphic to the $\Otwotwo$
action on $L^2(\BbbR^{2,2})$ known in this context \cite{howe-tan}. 
We give a brief review of the oscillator representation 
in appendix~\ref{app:oscrep}.

\subsection{Test states} 

Next, we seek a suitable linear space of test states
in~$\Haux$. The decomposition (\ref{comgen-rewrite}) suggests that we
make use of the eigenstates of the harmonic oscillator 
Hamiltonians~(\ref{harmoscuv-hams}). 
It is convenient 
to choose the eigenstates so that they are
also eigenstates of the 
angular momentum operators 
${\hat u}^1 {\hat p}^2 - {\hat u}^2
{\hat p}^1 = -i \partial_\alpha$ and 
${\hat v}^1 {\hat \pi}^2 - {\hat v}^2
{\hat \pi}^1 = -i \partial_\beta$. 
These eigenstates are 
\begin{equation}
\phi_{m,m';n,n'}
:= 
e^{i(m\alpha + m'\beta)}
\, 
u^{|m|} 
\, 
v^{|m'|} 
\, 
L^{|m|}_n ({\vec u}^2) 
\,
L^{|m'|}_{n'} ({\vec v}^2) 
\, 
\exp\left[-\casehalf({\vec u}^2 + {\vec v}^2) \right]
\ \ ,
\label{eigenstates}
\end{equation}
where the indices are integers with $n\ge0$ and $n'\ge0$, and the 
$L$'s are the generalized Laguerre polynomials 
\cite{arfken,magnusetal}. 
$\phi_{m,m';n,n'}$ is an eigenstate of ${\widehat H}^{\rm sho}_{\vec
  u}$ and ${\widehat H}^{\rm sho}_{\vec v}$ with the respective
eigenvalues $|m| + 2n$ and $|m'| + 2n'$, and it is 
an eigenstate of $-i \partial_\alpha$ and 
$-i \partial_\beta$ 
with the respective eigenvalues $m$ and~$m'$. 
The states $\phi_{m,m';n,n'}$
form a linearly independent and orthogonal set 
in~$\Haux$, satisfying 
\begin{equation}
\biglb( \phi_{m,m';n,n'} , 
\phi_{ {\tilde m}, {\tilde m}' ; {\tilde n} , {\tilde n}'}
\bigrb)_{\rm aux}
= 
\frac{\pi^2 (n + |m|)! \, (n' + |m'|)!}{n! \, (n')!}
\, 
\delta_{m, {\tilde m}} \, 
\delta_{m', {\tilde m}'} \, 
\delta_{n, {\tilde n}} \, 
\delta_{n', {\tilde n}'} 
\ \ , 
\label{phis-ip-aux}
\end{equation}
and their linear span 
${\tilde \Phi}$ 
is dense in~$\Haux$. 
${\tilde \Phi}$~consists of vectors of the form 
$P(\vec u, \vec v)
\exp\left[-\casehalf(\vec u^2 + \vec v^2) \right]$, where 
$P(\vec u, \vec v)$ is
an arbitrary polynomial in 
the four coordinates $(u^1,u^2,v^1,v^2)$: from this characterization
it is clear that ${\tilde \Phi}$ is mapped to itself by the
quantum constraint operators~(\ref{q-constraints}). Similarly,
recalling that the Algebraic Quantization
observable algebra $\Aphyplusstar$ is generated by 
(\ref{P-defs}) and 
the hatted counterparts
of~(\ref{perennials}), it is clear that ${\tilde \Phi}$ is mapped to
itself by~$\Aphyplusstar$. 

${\tilde \Phi}$ itself is not suitable for 
our RAQ the test state
space. First, there is a technical issue in that 
${\tilde \Phi}$ is not mapped to itself by the
$G$-action~$U$, as is immediate for example
from~(\ref{dilation-action}). 
The serious problem with 
${\tilde \Phi}$ is, 
however, that the group averaging integral
(\ref{GAM}) is not convergent, as we show in
appendix~\ref{app:convergence}: 
convergence fails when both angular momentum quantum numbers
vanish. We now show how to modify
${\tilde \Phi}$ so that the group averaging integral becomes
convergent, and we then use the group algebra technique of 
\cite{giumar-unique}
to generate a test state space that is invariant under $U$
and large enough for the uniqueness theorem of
\cite{giumar-unique} to apply. 

Let $\Phi_0$ be the linear span of the set
\begin{equation}
B_0 := 
\left\{ 
\left. \phi_{m,m';n,n'}
\, \vphantom{U_A^A}
\right| 
|m| + |m'|>0
\right\} \cup 
\left\{\left( \phi_{0,0;n,n'} + \phi_{0,0;n+1,n'+1} \right) 
\right\}
\ \ .
\label{Bnought-def}
\end{equation}
What motivates this definition is that $\Phi_0$ 
is mapped to itself by the 
Algebraic Quantization observable
algebra~$\Aphyplusstar$. 
To see this, recall from above that ${\tilde\Phi}$
is mapped to itself by~$\Aphyplusstar$. It is therefore sufficient 
to consider the situation in which an
element of $\Aphystar$ acts on a vector in $B_0$
and produces a vector whose expansion in the basis
$\left\{\phi_{m,m';n,n'} \right\}$ has components with 
$m=m'=0$. {}From~(\ref{P-defs}), 
(\ref{polartaus}), 
and the angle
dependence in $\phi_{m,m';n,n'}$~(\ref{eigenstates}), 
we see that the only nontrivial instance of how this can happen 
is the action of 
${\widehat\tau}_\pm^\eta$ on $\phi_{\mp1,\mp\eta;n,n'}$, which 
reads by explicit computation \cite{magnusetal-recurrence} 
\begin{equation}
{\widehat\tau}_\pm^\eta \phi_{\mp1,\mp\eta;n,n'} 
= 
(n+1)(n'+1) 
\left( \phi_{0,0;n,n'} + \phi_{0,0;n+1,n'+1} \right) 
\ \ , 
\label{taus-to-zero}
\end{equation}
and this is in the linear span of~$B_0$. 
Thus $\Phi_0$ is mapped to itself by~$\Aphyplusstar$. 

We claim that $\Phi_0$ is dense in~$\Haux$. 
To show this, recall from above that 
$\left\{\phi_{m,m';n,n'}
\right\}$ is an orthogonal Hilbert space basis for~$\Haux$.  
It is therefore sufficient
to show that the 
linear subspace $W \subset \Phi$ spanned by 
$\left\{ \left( \phi_{0,0;n,n'} + \phi_{0,0;n+1,n'+1} \right) 
\right\}$ is dense in the Hilbert subspace 
${\cal H}_0 \subset \Haux$ spanned by 
$\left\{\phi_{0,0;n,n'} \right\}$. Suppose this is false. Then there
exists a nonzero vector $v \in {\cal H}_0$ that is in the orthogonal
complement of the closure of~$W$. 
As $v \in {\cal H}_0$, we can write 
$v = \sum_{n,n'} b_{n,n'} \phi_{0,0;n,n'}$, 
where the coefficients satisfy 
$\sum_{n,n'} |b_{n,n'}|^2 < \infty$ by~(\ref{phis-ip-aux}), 
and at least one coefficient is nonzero. However, 
the orthogonality of $v$ 
with each $\left( \phi_{0,0;n,n'} + \phi_{0,0;n+1,n'+1}
\right)\in W$ implies $b_{n,n'} = - b_{n+1,n'+1}$ for all $n$
and~$n'$, and the sum $\sum_{n,n'} |b_{n,n'}|^2$ therefore diverges, 
which is a contradiction. 
Thus $\Phi_0$ is dense in~$\Haux$. 

The crucial property of $\Phi_0$ is
that the group averaging integral
(\ref{GAM}) converges in absolute value for all $\phi_1$ and
$\phi_2$ in $\Phi_0$. This is shown in 
appendix~\ref{app:convergence}. 

As $\Phi_0$ is not mapped to itself by~$U$, $\Phi_0$ does not
technically qualify as a test state space in our version of RAQ\null.
A~simple remedy would be to consider the space~$\Phi_0'$, which is the
closure of $\Phi_0$ under the algebra generated by the operators
$U(g)$ for $g\in G$.  $\Phi_0'$~is clearly dense in $\Haux$ and
invariant under~$U$, and it thus satisfies the RAQ test state space
conditions, and one could indeed successfully complete RAQ with
$\Phi_0'$ as the test state space. However, we wish to work with a test
state space to which the uniqueness theorem of Giulini and Marolf
\cite{giumar-unique} applies.  To this end, recall from section
\ref{sec:RAQoutline} that an $L^1$ function $h$ on $G$ defines on
$\Haux$ the bounded operator ${\hat{h}} := \int_G dg \, h(g) U(g)$,
and the set of all such operators forms an algebra~$\Agrouphat$.  Let
now $\Phi$ be the closure of $\Phi_0'$ under the action
of~$\Agrouphat$. It is clear that $\Phi$ is dense in $\Haux$ and
invariant under~$U$, and $\Phi$ thus satisfies the RAQ test state
space conditions.
It is
also clear that $\Phi$ is mapped to itself by~$\Agrouphat$, while
$\Phi_0'$ is not.

We now adopt $\Phi$ as the RAQ test state space. As $\Phi_0$ is
mapped to itself by~$\Aphyplusstar$, so is~$\Phi$, and the 
RAQ observable algebra $\Aobs$
therefore contains $\Aphyplusstar$ as a subalgebra.

As a final remark, we note that $\Phi_0$ is mapped to itself by the
quantum constraint operators (\ref{q-constraints})
\cite{magnusetal-recurrence}, and therefore $\Phi_0'$ and $\Phi$ are
also mapped to themselves by these operators.  $\Phi_0$, $\Phi_0'$,
and $\Phi$ would therefore all qualify as test state spaces in
formulations of RAQ that solve the constraints in terms of the
constraint operators rather than in terms of the $G$-action~$U$
\cite{epistle,giumar-general}.

\subsection{Group averaging and the physical Hilbert space} 

Consider now the group averaging. As mentioned above, 
the integral in (\ref{GAM}) 
converges in absolute value for all $\phi_1$ and $\phi_2$
in~$\Phi_0$. It follows from Lemma 2 in \cite{giumar-unique} that 
the integral in (\ref{GAM}) converges in absolute value for 
all $\phi_1$ and $\phi_2$ in~$\Phi$. 
The map $\eta$ is therefore well defined by (\ref{GAM})
and~(\ref{GAM-eta}), and it satisfies the rigging map postulates
with the possible exception of positivity. 

To evaluate~$\eta$, 
let $\phi_i\in\Phi$, and let $h_i$ be $L^1$
functions on~$G$. We then have from (\ref{GAM}) and (\ref{GAM-eta}) 
\cite{giumar-unique}
\begin{mathletters}
\begin{eqnarray}
\eta( {\hat{h}}_1 \phi_1) [\phi_2] 
&=& 
\overline{
\left(\int_G d g \, h_1(g) \right)
} \, 
\eta(\phi_1)[\phi_2] 
\ \ ,
\\
\eta( \phi_1) [ {\hat{h}}_2 \phi_2] 
&=& 
\left(\int_G d g \, h_2(g) \right)
\eta(\phi_1)[\phi_2] 
\ \ .
\end{eqnarray}
\end{mathletters}
As further 
$\eta(\phi_1) [ U(g_0) \phi_2]  = \eta\biglb(U(g_0)\phi_1\bigrb) [\phi_2] 
= \eta(\phi_1) [\phi_2]$, it suffices to evaluate 
$\eta(\phi_1) [\phi_2]$ for $\phi_1$ and $\phi_2$ in the set
$B_0$~(\ref{Bnought-def}). 

The explicit evaluation of 
$\eta$ is done in appendix~\ref{app:rigg-ev}. 
We can represent the vectors in the image of 
$\eta$ as functions on $\BbbR^4 = \{(\vec u, \vec
v)\}$, acting on the test states $\phi\in\Phi$ by 
\begin{equation}
f[\phi] = \int d^2 \vec u \, d^2 \vec v \, 
f (\vec u, \vec v) \phi(\vec u, \vec v)
\ \ . 
\label{dual-action}
\end{equation}
We find 
\begin{mathletters}
\label{aver-results}
\begin{eqnarray}
&&
\eta (\phi_{m,m';n,n'}) 
= 
2\pi^2 {(-1)}^{n} 
{[{\rm sgn}( m)]}^{m}
\delta_{|m|, |m'|} \, \delta_{n, n'}
\, 
\frac{(n+|m|)!}{|m| \, n!}
\, 
f_{m,(m'/m)}
\ \ ,
\ \ |m| + |m'|>0
\ \ , 
\nonumber
\\
&&
\label{aver-result-nonzero}
\\
&&
\eta 
\left(
\phi_{0,0;n,n'} + \phi_{0,0;n+1,n'+1}
\right) 
= 
0 
\ \ , 
\label{aver-result-zero}
\end{eqnarray}
\end{mathletters}
where the functions $f_{m,\epsilon}$, 
with  $m\in\BbbZ\setminus\{0\}$ and $\epsilon \in \{1,-1\}$, 
are defined by 
\begin{equation}
f_{m,\epsilon} := 
J_{m} (uv) \, e^{-i m(\alpha + \epsilon \beta)}
\ \ . 
\label{eta-imagebasis}
\end{equation}
The action (\ref{dual-action})
of $f_{m,\epsilon}$ on the vectors in $B_0$ reads 
\cite{magnusetal-integrals} 
\begin{mathletters}
\label{fs-on-phis}
\begin{eqnarray}
&&
f_{m,\epsilon}[\phi_{{\tilde m},{\tilde m}';n,n'}]
= 
2\pi^2 {(-1)}^{n} 
{[{\rm sgn}( m)]}^{m}
\delta_{m, {\tilde m}} \, 
\delta_{\epsilon m, {\tilde m}'} \, 
\delta_{n, n'}
\, 
\frac{(n+|m|)!}{n!}
\ \ ,
\ \ 
|{\tilde m}| + |{\tilde m}'|>0
\ \ ,
\label{fs-on-phis-nonzero}
\\
&&
f_{m,\epsilon}
[ \phi_{0,0;n,n'} + \phi_{0,0;n+1,n'+1} ]
= 0 
\ \ .
\end{eqnarray}
\end{mathletters}
{}From this it is clear that 
the set $\left\{f_{m,\epsilon} \mid 
  m\in \BbbZ\setminus\{0\}, \epsilon = \pm1  \right\}$ 
is linearly independent in $\Phi^*$ 
and a basis for the image of~$\eta$. 

What remains is to evaluate the 
(prospective) inner product on the image of~$\eta$. 
{}From~(\ref{phys-ip}), 
(\ref{aver-result-nonzero}), and~(\ref{fs-on-phis-nonzero}),  we find
\begin{equation}
\bigl(f_{m,\epsilon} , f_{m',\epsilon'} \bigr)_{\rm phys} 
= 
|m| \, 
\delta_{m, m'} \delta_{\epsilon, \epsilon'}
\ \ . 
\label{phys-ip-basis}
\end{equation}
As (\ref{phys-ip-basis}) is positive
definite, all the rigging map postulates are satisfied, and
(\ref{phys-ip-basis}) does define an 
inner product on the image of~$\eta$. 
The physical Hilbert space $\Hphys$ is obtained by completion. 
The asymptotic large order expansion of $J_m$ \cite{abra-stegun} shows 
that every vector in $\Hphys$ can be represented
as a function on 
$\BbbR^4 = \{(\vec u, \vec v)\}$. 

Finally, as $\Phi$ is invariant under~$\Agrouphat$, the
assumptions of the uniqueness theorem of Giulini and Marolf
are satisfied. It follows that every rigging map for our triple 
$(\Haux,U,\Phi)$ is a multiple of the group averaging rigging 
map~$\eta$.

\subsection{Observables 
and the relation to Algebraic Quantization} 

As we have emphasized, the RAQ observable algebra $\Aobs$ 
contains the Algebraic Quantization observable algebra
$\Aphyplusstar$ as a subalgebra, and the star-operation on 
$\Aphyplusstar$ is
the adjoint map of~$\Haux$. It follows that the antilinear
representation of $\Aobs$ on $\Hphys$ contains 
an antilinear representation $\rho_+$ of~$\Aphyplusstar$, 
and in $\rho_+$ the star-operation on 
$\Aphyplusstar$
is the adjoint map of~$\Hphys$. $\rho_+$~acts 
on the basis 
$\left\{f_{m,\epsilon} \mid
  m\in \BbbZ\setminus\{0\}, \epsilon=\pm1 \right\}$ 
of $\Hphys$ as 
\begin{mathletters}
\label{astarantirep}
\begin{eqnarray}
\rho_+({\widehat\tau}_0^\eta):
&& 
f_{m,\epsilon} 
\mapsto 
\delta^{\eta,\epsilon} m 
f_{m,\epsilon} 
\ \ ,
\\
\rho_+({\widehat\tau}_\pm^\eta):  
&&
f_{m,\epsilon} 
\mapsto \delta^{\eta,\epsilon} m 
f_{m\pm1,\epsilon} 
\ \ , 
\\
\rho_+(
{\widehat P}_{\epsilon_1, \epsilon_2}): 
&&
f_{m,\epsilon} 
\mapsto
f_{\epsilon_1 m,\epsilon_2 \epsilon} 
\ \ ,
\end{eqnarray}
\end{mathletters}
where $f_{0,\epsilon}$, whenever it appears on the right-hand side, is
understood to mean zero. 

Comparing (\ref{astarantirep})
to (\ref{astarrepV}) and~(\ref{P-repV}), 
and the RAQ inner product (\ref{phys-ip-basis})
to the Algebraic Quantization inner product~(\ref{norm-on-Vs}), we see
that $\rho_+$ is anti-isomorphic to the representation of 
$\Aphyplusstar$ 
on the Hilbert space ${\cal H}_+$ obtained 
in the Algebraic Quantization of
section~\ref{sec:oldAQ}, 
provided the inner products are normalized to agree. 
The $\Otwotwo$ action on ${\cal H}_+$ found in section \ref{sec:oldAQ}
is anti-isomorphic to the $\Otwotwo$-action on $\Hphys$ 
induced by the $\Otwotwo$ action on~$\Haux$. 
In this sense, the 
RAQ quantum theory
contains the Algebraic Quantization quantum theory.

\section{Discussion}
\label{sec:discussion}

In this paper we have compared the Algebraic Quantization (AQ)
framework and the Refined Algebraic Quantization (RAQ) framework in a
constrained Hamiltonian system with unreduced phase space
$\Gamma=T^*\BbbR^4$ and gauge group~$\SLtwor$. In both approaches we
used input motivated by the structure of the classical constraints as
quadratic functions on~$\Gamma$. 
In~AQ, we first solved the
constraints on a suitable vector space, promoted an explicitly-known
classical observable algebra into the quantum operator
star-algebra~$\Aphyplusstar$, and determined the inner product by
requiring the star-operation on $\Aphyplusstar$ to coincide with the
adjoint operation.  In~RAQ, we chose the auxiliary Hilbert space
$\Haux$ to be $L^2$ over the unreduced configuration space~$\BbbR^4$,
and we promoted the classical $\SLtwor$ gauge transformations on
$\Gamma$ into a unitary $\SLtwor$-action on~$\Haux$.
We
took particular care to choose the RAQ test state space
$\Phi\subset\Haux$ so
that the RAQ observable algebra $\Aobs$ contains~$\Aphyplusstar$.
Considering the similarity in these inputs, it is not surprising that
the RAQ quantum theory turned out to contain the AQ quantum theory. We
also investigated the $\Otwotwo$ group actions underlying the 
classical and quantum observable algebras, 
and we showed that the AQ quantum theory can be
recovered through Isham's group theoretic quantization
framework. 

Both AQ and RAQ encountered with the zero angular momentum states 
a technical difficulty whose origin is in
the structure of a certain
pathological subset of the classical reduced phase space. 
The remedy was to ensure that
such states do not appear in the physical Hilbert space. 
In~AQ, the problem appeared in the guise of ``zero norm''
states in the prospective Hilbert space, and the cure was simply to
drop the states already from the vector space on which the
constraints are solved. In~RAQ, on the other hand, the problem
appeared as the divergence of the 
group averaging, and the cure now was to
modify the space of test states. However, as the RAQ observable
algebra is defined in terms of the test state space, 
the modification needed to be quite subtle 
in order that the 
RAQ observable algebra could still be meaningfully compared
with the AQ observable algebra: here we took
advantage of the
explicit knowledge of the operators in~$\Aphyplusstar$. 
This illustrates well how neither AQ nor RAQ is a {\em prescription\/} for
quantization: they are schemes that need input at various
steps, and making successful choices in the `early' steps may require
hindsight from the `later' steps
\cite{ashtate,giumar-general,tate-remark}. 
Also, this illustrates that although RAQ does not assume a single
observable to be explicitly constructed, the knowledge of some
observables of interest can be quite useful in making good choices 
at the various steps of~RAQ\null. 

As discussed in \cite{monrovthie}, the constraint algebra of our
system is analogous to the constraint algebra of general relativity. 
Among the three constraints~(\ref{constraints}), $H_1$~and 
$H_2$ are ``Hamiltonian''-type, quadratic in the momenta, 
while $D$ is 
``momentum''-type, linear in the momenta, and the mixing of these two
types of constraints in (\ref{PBalg-constraints}) is as in 
general relativity \cite{teitel-gralgebra}. 
One consequence of this analogy 
is that one could introduce and investigate 
in our system also group
averaging with Teitelboim's ``causal'' boundary 
condition
\cite{teitel-tril1,teitel-tril2,henn-teitel-particle}. In general
relativity, this condition proposes 
that only positive lapses contribute to
the path integral that defines the quantum mechanical propagation
amplitude. If $H_1$ and $H_2$ are adopted as the analogue of the
Hamiltonian constraint of general relativity at two spatial
points \cite{monrovthie}, the causal boundary condition in our system
yields an average over the semigroup of 
$\SLtwor$ matrices whose all entries are positive: 
integrating first over the lapses and then over the shift, the 
$\SLtwor$ elements emerge from the amplitude folding of 
\cite{teitel-tril1,teitel-tril2} in the form 
\begin{equation}
\exp(\nu h) \exp(\nu_+ e^+ + \nu_- e^-)
\ \ , 
\label{teitel-elements}
\end{equation}
where $-\infty < \nu <\infty$ and $\nu_\pm>0$, 
and we have explicitly verified that the measure emerging from the
ghost integrations of 
\cite{teitel-tril1,teitel-tril2} is the 
$\SLtwor$ 
Haar measure in the parametrization~(\ref{teitel-elements}). 
It might be interesting to see whether a scattering theory of the type 
considered in \cite{teitel-tril1,teitel-tril2,henn-teitel-particle}
could be built on the causal boundary condition in our system. 

We note in this context that 
allowing 
$\nu$ and $\nu_\pm$ to take all real values in 
(\ref{teitel-elements}) does not cover all of $\SLtwor$, and in
particular it does not reach those matrices where the product of the
diagonal elements is negative. 
In our system, the decomposition of the
quantum propagation amplitude in the form given in 
\cite{teitel-tril1,teitel-tril2}, first integrating over the lapses and 
then over the shift, does thus not yield an average over the
whole group when the lapses and the shift are allowed to take
all real values. This phenomenon occurs also upon
considering 
(\ref{teitel-elements}) in the group
$\PSLtwor = \SLtwor/\{\pm\openone\} \simeq \Octwoone$. The phenomenon
is therefore not a consequence of
the fact that the exponential map from 
$\sltwor$ to $\SLtwor$ is not onto, as the exponential map from 
$\sltwor$ to $\PSLtwor$ is.

We saw in section 
\ref{sec:classical} that the $\Otwotwo$ action on
$\Gamma$ looks simple in the polarization in which 
$(\vec u, \vec \pi)$ are the `coordinates' and 
$(\vec p, -\vec v)$ are the `momenta'. Similarly, we noted in section 
\ref{sec:RAQus} that 
the $\Otwotwo$ action on $\Haux = L^2 \biglb(\{(\vec u,
\vec v)\}\bigrb)$ looks simple when Fourier-transformed to 
$L^2 \biglb(\{\vec u, \vec \pi)\}\bigrb)$. 
Attempting to quantize the system in a $(\vec u, \vec
\pi)$-representation would however present difficulties. 
Adopting the $(\vec u, \vec
\pi)$-representation in Algebraic Quantization and proceeding as in 
section~\ref{sec:oldAQ}, one finds that 
the constraints cannot be solved in terms of 
smooth functions: the constraint 
${\widehat H}_2\Psi=0$ implies that the support of $\Psi(\vec u, \vec
\pi)$ would need
to be in some sense 
at ${\vec\pi}^2 - {\vec u}^2=0$. 
The reason underlying this difficulty is
precisely that our solutions to the constraints in the $(\vec u, \vec
v)$-representation are not square integrable, or even integrable, and
Fourier-transforming them to a $(\vec u, \vec
\pi)$-representation is a priori not defined. 
In~RAQ, in contrast, the Fourier-transform to the $(\vec u, \vec
\pi)$-representation is well-defined in~$\Haux$, there is no
obstacle to constructing in this representation~$\Phi$, 
the $G$-action, or the group
averaging sesquilinear form~(\ref{GAM}), and proving the
absolute convergence of the integral in (\ref{GAM}) is in fact
technically simpler than in the 
$(\vec u, \vec v)$-representation. At the abstract level, one
thus recovers isomorphic RAQ quantum theories in the 
$(\vec u, \vec v)$-representation and the 
$(\vec u, \vec \pi)$-representation. 
The difficulty of doing RAQ in the 
$(\vec u, \vec \pi)$-representation is a more 
practical one, namely, 
that the methods of appendix \ref{app:rigg-ev}
now do not yield a representation of the image of
$\eta$ as functions on 
$\BbbR^4 = \{(\vec u, \vec \pi)\}$, 
and one needs some other way to prove that
$\eta$ is positive and to evaluate $\eta$ in some practical
fashion. 

The classical system admits a generalization in which 
$\vec u$ and $\vec v$ in the action functional 
(\ref{action})--(\ref{constraints}) 
have respectively $r$ and $s$
components, for any nonnegative integers $r$ and~$s$. 
The phase
space is $\Gamma_{r,s} := T^*\BbbR^{r,s}$, the gauge group
generated by the constraints is 
still $\SLtwor$, and 
$\Gamma_{r,s}$ has a natural $\Ogrouprs$-action that commutes with the 
$\SLtwor$-action. 
One expects that this generalized system could 
be quantized with our
methods, and that the quantum theory would reflect properties 
of the oscillator representation 
on $L^2(\BbbR^{r,s})$ \cite{howe-tan,howe-decomp,lands-pairs}. 
It is also possible to generalize the system to 
certain other gauge groups of interest by
minor modification of the constraint structure in
(\ref{action})--(\ref{constraints}), such as to the
$(1+1)$ Poincare group, or to the affine group on $\BbbR$ (which is
nonunimodular). 
We leave such generalizations subject to future work. 

{\em Note added:\/} 
After this work was completed, 
a quantization of the system in the 
Algebraic
Constraint Quantization framework of 
\cite{trunk-kepler,trunk-pseudorigid} was posted 
in 
\cite{trunk-sltwor}. 
As noted in \cite{trunk-sltwor}, the quantum theory recovered therein
is in essence identical to our 
Algebraic Quantization quantum theory.


\acknowledgments
We thank 
Nico Giulini, 
Don Marolf, 
Merced Montesinos, 
Alan Rendall, 
and 
Thomas Thiemann
for helpful discussions and correspondence, 
Michael Trunk for bringing \cite{trunk-sltwor} to our attention, 
and 
Martin Bojowald, 
Hans Kastrup, 
and 
Thomas Strobl 
for discussions
on \cite{bojo-etal,bojo-strobl}. 
J.L.~thanks John Friedman for his hospitality at the
University of Wisconsin-Milwaukee, where part of this work was done. 
This work was supported in part by NSF Grant PHY-9900791.


\appendix
\section{Iwasawa decomposition of $\SLtwor$}
\label{app:iwasawa}

In this appendix we collect some well-known properties 
of $\SLtwor$. The notation follows \cite{howe-tan}. 

$\SLtwor$ consists of real $2 \times 2$ matrices with unit
determinant, 
\begin{equation}
g = \pmatrix{a & b \cr c & d \cr}
\ \ , \ \ ad - bc =1
\ \ .
\label{SLtwomatrix}
\end{equation}
Each element of $\SLtwor$ admits a unique Iwasawa decomposition, 
\begin{equation}
g = 
\pmatrix{1 & 0 \cr \mu & 1 \cr}
\pmatrix{e^\lambda & 0 \cr 0 & e^{-\lambda} \cr}
\pmatrix{\cos \theta & \sin \theta \cr 
- \sin \theta & \cos \theta \cr}
\ \ ,
\label{iwasawa}
\end{equation}
where $\mu\in\BbbR$, $\lambda\in\BbbR$, and 
$0\le\theta<2\pi$. 
In terms of the parametrization~(\ref{iwasawa}), 
the left and right invariant Haar
measure reads $e^{2\lambda}\,d\lambda\,d\mu\,d\theta$. 

A~standard basis for the 
Lie algebra $\sltwor$ consists of the three
matrices 
\begin{eqnarray}
h &:=& \pmatrix{1 & 0 \cr 0 & -1 \cr}
\ \ ,
\nonumber
\\
e^+ &:=& \pmatrix{0 & 1 \cr 0 & 0 \cr}
\ \ ,
\nonumber
\\
e^- &:=& \pmatrix{0 & 0 \cr 1 & 0 \cr}
\ \ ,
\label{sltwo-basis}
\end{eqnarray}
whose commutators are 
\begin{eqnarray}
\left[ h \, , \,  e^+  \right] 
& = & 
2e^+
\ \ ,
\nonumber
\\
\left[ h \, , \,  e^-  \right] 
& = & 
-2e^-
\ \ ,
\nonumber
\\
\left[ e^+ \, , \, e^- \right]
& = & 
h 
\ \ .
\label{sltwo-basiscomm}
\end{eqnarray}
A second standard basis is 
\begin{eqnarray}
&&\gamma_0 :=
\casehalf (e^+ - e^-)
\ \ ,
\nonumber
\\
&&\gamma_1 := 
\casehalf (e^+ + e^-)
\ \ ,
\nonumber
\\
&&\gamma_2 := 
\casehalf h 
\ \ ,
\label{sltwo-basis2}
\end{eqnarray}
with the commutators 
\begin{eqnarray}
\left[ \gamma_1 \, , \, \gamma_2 \right]
& = & 
- \gamma_0
\ \ ,
\nonumber
\\
\left[ \gamma_2 \, , \, \gamma_0 \right] 
& = & 
\gamma_1
\ \ ,
\nonumber
\\
\left[ \gamma_0 \, , \, \gamma_1 \right]
& = & 
\gamma_2
\ \ .
\label{sltwo-basiscomm2}
\end{eqnarray}

Each of the three matrices in (\ref{iwasawa}) is in the image of the
exponential map from $\sltwor$ to $\SLtwor$. 
In terms of the exponential map, 
(\ref{iwasawa}) reads 
\begin{equation}
g = 
\exp(\mu e^-)
\exp(\lambda h)
\exp[\theta(e^+ - e^-)]
\ \ . 
\label{iwa-abs}
\end{equation}

The decomposition (\ref{iwasawa}) encodes the first homotopy group
$\BbbZ$ of $\SLtwor$ entirely in the rightmost factor. The 
quotient map from $\SLtwor$ to $\PSLtwor = \SLtwor/\{\pm\openone\}
\simeq \Octwoone$ [the connected component
of $\Otwoone$] acts in the
decomposition (\ref{iwasawa}) by the identification
$(\mu,\lambda,\theta) \sim 
(\mu,\lambda,\theta+\pi)$. 
A~unique Iwasawa decomposition
of the form (\ref{iwa-abs}) holds therefore also for covering groups
of $\Octwoone$: for the $n$-fold covering $0\le \theta < n\pi$, and
for the universal covering $-\infty < \theta < \infty$.

\section{Oscillator representation of the double cover of
  $\SLtwor$} 
\label{app:oscrep}

In this appendix we recall some properties of the
oscillator representation of the double cover
of $\SLtwor$ \cite{howe-tan}. 
We denote in this appendix the double cover of 
$\SLtwor$ by $\SLtwordouble$.

\subsection{Oscillator representation on $L^2(\BbbR)$} 

Consider on $L^2(\BbbR)$ the three 
essentially self-adjoint operators 
\begin{mathletters}
\label{osc-constraints}
\begin{eqnarray}
{\widehat H}_1 & := & 
- \casehalf \partial_x^2
\ \ ,
\\
{\widehat H}_2 & := & 
- \casehalf x^2 
\ \ ,
\\
{\widehat D} & := & 
-i 
\left( 
x \partial_x + \casehalf \right) 
\ \ , 
\end{eqnarray}
\end{mathletters}
whose commutators form the $\sltwor$ algebra~(\ref{q-constraints-alg}). 
Exponentiation yields a unitary 
representation $\omega$ 
of the universal covering group of $\SLtwor$ on $L^2(\BbbR)$. 
The group elements that appear in the 
Iwasawa decomposition (\ref{iwa-abs}) are represented by 
\begin{mathletters}
\label{osc-uss}
\begin{eqnarray}
\omega \biglb( \exp (\mu e^- ) \bigrb) 
&=& 
\exp (-i \mu {\widehat H}_2 )
\ \ ,
\\
\omega  \biglb( \exp(\lambda h) \bigrb) 
&=& 
\exp \biglb( -i \lambda {\widehat D} \bigrb) 
\ \ ,
\\
\omega \biglb( \exp[\theta(e^+ - e^-)] \bigrb) 
&=& 
\exp \biglb(-i \theta ( {\widehat H}_1 - {\widehat H}_2) \bigrb)
\ \ . 
\end{eqnarray}
\end{mathletters} 
The two first operators in (\ref{osc-uss}) act on functions 
$\psi(x)$ as 
\begin{mathletters}
\begin{eqnarray}
{[ \exp (-i \mu {\widehat H}_2 ) \, \psi ](x)}
&=& 
e^{i \mu  x^2/2} \, \psi(x)
\ \ ,
\\
{[ \exp (-i \lambda {\widehat D}) \, \psi ](x)}
&=& 
e^{-\lambda/2} \, \psi(e^{-\lambda}x)
\ \ , 
\end{eqnarray}
\end{mathletters}
while $\exp \biglb(-i \theta ( {\widehat H}_1 - {\widehat H}_2)
\bigrb)$ is the unit mass and frequency 
harmonic oscillator evolution operator. As 
$\exp \biglb(-i \theta ( {\widehat H}_1 - {\widehat H}_2)
\bigrb)$ is periodic in $\theta$ with period~$4\pi$, 
$\omega$ is a representation of $\SLtwordouble$ 
but not a representation of $\SLtwor$. 


It is evident that $\omega$ decomposes into a sum of two
unitary representations, one acting on even and the other on odd
functions. 
It can be shown that these two representations are
irreducible \cite{howe-tan}.

The oscillator representation can be formally written as 
\begin{equation}
[\omega(g) \, \psi] (x)
= 
\int
\frac{dy}{\sqrt{2\pi i b}} 
\exp {\left[
\frac{i
\left(
ay^2 + dx^2  - 2 x y
\right)}
{2b} 
\right] } 
\, \psi(y)
\ \ ,
\label{gen-osc-rep}
\end{equation}
where $a$, $b$ and $d$ are as shown in (\ref{SLtwomatrix}) in the 
$\SLtwor$ representative of $g$ [while $g$ itself is 
in $\SLtwordouble$]. 
The singularities and
branch cuts in the integral kernel in 
(\ref{gen-osc-rep})
must 
however be interpreted consistently with the 
unambiguously-defined left-hand side. For example, when $g =
\exp[\theta(e^+ - e^-)]$, 
$\omega(g)$ is the harmonic oscillator evolution operator, for
which
$a = d = \cos\theta$ and $b=\sin\theta$, 
and the integral kernel in
(\ref{gen-osc-rep}) is singular at 
$\theta=\pi n$. 

The integral kernel in (\ref{gen-osc-rep}) can
be derived from the $\SLtwor$ action that the classical counterparts
of the operators (\ref{osc-constraints}) generate on~$T^*\BbbR$.
Writing $g \in \SLtwor$ as in~(\ref{SLtwomatrix}), and denoting by
$(q,p)$ the usual canonical chart on~$T^*\BbbR$, this action reads
\begin{equation}
\pmatrix{q \cr p \cr}
\mapsto 
\pmatrix{q' \cr p' \cr}
= 
\pmatrix{a & b \cr c & d \cr}
\pmatrix{q \cr p \cr}
\ \ . 
\label{class-oscaction}
\end{equation}
(\ref{class-oscaction})~preserves 
the symplectic structure $dp \wedge dq$ 
and is therefore a canonical transformation. 
For $b\ne0$, one can
express the old and new momenta as functions of the old and new
coordinates, and the canonical transformation has 
then a generating
function $S(q,q')$, satisfying 
\begin{equation}
p'(q,q')\,  dq' - p (q,q') \, dq = dS(q,q')
\ \ . 
\end{equation}
Simple algebra yields
\begin{equation}
S(q,q')= \frac{aq^2+dq'{}^2-2qq'}{2b}
\ \ .
\label{cgener-function} 
\end{equation}
As $S(q,q')$ (\ref{cgener-function})
is quadratic in $q$ and~$q'$,  
the integral kernel of the corresponding
unitary transformation consists of the exponential $\exp[i
S(q,q')]$ and a prefactor that does not depend on $q$
or~$q'$. Imposing unitarity yields the prefactor 
shown in~(\ref{gen-osc-rep}).

\subsection{Oscillator representation on $L^2(\BbbR^{r,s})$} 

Inverting the signs of both $e^+$ and $e^-$ in the basis 
(\ref{sltwo-basis}) of $\sltwor$ is an automorphism 
of~$\sltwor$. Inverting the signs of 
${\widehat H}_1$ and ${\widehat H}_2$ in 
(\ref{osc-constraints}) and proceeding as above yields therefore
a representation $\omega^*$ of 
$\SLtwordouble$ on~$L^2(\BbbR)$. 
The tensor product $\omega_{r,s}$ of $r$ copies of $\omega$ and $s$
copies of $\omega^*$ is naturally realized as a representation of 
$\SLtwordouble$ on 
$L^2(\BbbR^{r,s})$, each $\omega$ acting on one of the first $r$
coordinates and each $\omega^*$ acting on one of the last $s$
coordinates. 
$\omega_{r,s}$~is a representation of $\SLtwor$ iff $r+s$
is even. 

The group $\Ogrouprs$ has a natural action on $L^2(\BbbR^{r,s})$ by
$\psi(x) \mapsto \psi(a^{-1}x)$, where $a$ is in the defining matrix
representation of~$\Ogrouprs$. 
This $\Ogrouprs$ action commutes with~$\omega_{r,s}$, 
and the spectral decomposition of one completely
determines the spectral decomposition of the other
\cite{howe-tan,howe-decomp}. 

The representation $U$ of $\SLtwor$ on $\Haux$ 
introduced in 
section \ref{sec:RAQus}, and the representation of $\Otwotwo$ on 
$\Haux$ generated by the observable algebra $\Aphyplusstar$ therein, 
are
isomorphic to the above structure with $r=s=2$. The
isomorphism is the Fourier transform in the last two coordinates in 
$\Haux \simeq L^2(\BbbR^{2,2})$.

\section{Convergence of the group averaging}
\label{app:convergence}

In this appendix we show that the integral in~(\ref{GAM}), 
\begin{equation}
\int_G d g \, \biglb(\phi_2 , U(g) \phi_1 \bigrb)_{\rm aux}
\ \ ,
\label{GAMtoo}
\end{equation}
converges in absolute value for all 
$\phi_1$ and $\phi_2$ in the 
space $\Phi_0$ defined in section~\ref{sec:RAQus}. 

It suffices to consider $\phi_1$ and $\phi_2$ 
in the set $B_0$~(\ref{Bnought-def}).  
As the 
operators
$-i\partial_\alpha$ and $-i\partial_\beta$ (which belong to
$\Aobs$) commute with~$U(g)$, 
it suffices to consider the case where 
$\phi_1$ and $\phi_2$ have the same angular momentum quantum numbers,
for otherwise the integrand in (\ref{GAMtoo}) 
vanishes by the orthogonality~(\ref{phis-ip-aux}). 

We now consider separately the case where at least one angular
momentum is nonzero and the case where both angular momenta are zero.

\subsection{At least one angular momentum nonzero}

We set $\phi_1 = \phi_{m,m';n,n'}$ and 
$\phi_2 =  \phi_{ m, m' ; {\tilde n} , {\tilde
    n}'}$, where 
$|m|+|m'|>0$. 

We write $g$ in the Iwasawa decomposition~(\ref{iwasawa}). 
By~(\ref{Us-iwa}), 
$U(g)$ is given by 
\begin{equation}
U(g) = 
\exp (-i \mu {\widehat H}_2 )
\exp \biglb( -i \lambda {\widehat D} \bigrb) 
\exp \biglb(-i \theta ( {\widehat H}_1 - {\widehat H}_2) \bigrb)
\ \ . 
\label{app-rep-iwasawa}
\end{equation}
The Haar measure in (\ref{GAMtoo}) reads 
$dg = e^{2\lambda}\,d\lambda\,d\mu\,d\theta$, and the integration
is over all real values of 
$\lambda$ and $\mu$ and over one $2\pi$ cycle 
in~$\theta$. As 
$\phi_1$ is an eigenstate
of the rightmost operator in 
(\ref{app-rep-iwasawa}) with an eigenvalue of absolute value~$1$, it 
suffices to set $\theta=0$ and consider 
the integral over $\lambda$ and $\mu$ in the measure 
$e^{2\lambda}\,d\lambda\,d\mu$. 

Let thus $U(g)$ be as in (\ref{app-rep-iwasawa}) 
with $\theta=0$ and $\mu\neq0$. By~(\ref{nullanddila-action}), 
we have 
\begin{eqnarray}
\biglb(\phi_2 ,  U(g) && \phi_1  \bigrb)_{\rm aux} 
= 
\frac{4 \pi^2 [{\rm sgn}(m')]^{m'} z^{(k'-k)/2}}
{i^{(m'+1)} \mu}
\nonumber
\\
\times 
\int 
&&
du \, dv \, dv'
\> 
u^{2k+1}
v^{k'+1}
{(v')}^{k'+1}
J_{k'}(v v'/\mu) 
L^k_{\tilde n}(u^2) 
L^{k'}_{{\tilde n}'}(v^2) 
L^k_n(u^2/z) 
L^{k'}_{n'} \biglb(z {(v')}^2\bigrb) 
\nonumber
\\
&&
\times
\exp \left\{
- \casehalf 
\left[ 1 + (1/z) - i \mu \right] u^2 
- \casehalf 
\left[ 1  - (i/\mu) \right] v^2 
- \casehalf 
\left[ z  - (i/\mu) \right] {(v')}^2 
\right\}
\ \ ,
\label{element-2}
\end{eqnarray}
where 
$k := |m|$, 
$k' := |m'|$, 
$z:=e^{2\lambda}$, 
and the integration is over positive values of 
$u$, $v$, and~$v'$. 
The Bessel function $J_{k'}(v v'/\mu)$ has emerged from
performing the angular part of the $d^2 \vec v'$ 
integral in~(\ref{null-action}). 
Here, and from now on, the individual components of $\vec u$ and $\vec 
v$ will not appear, and we always write $u = \sqrt{{\vec u}^2}$, 
$u^2 := {\vec u}^2$, and so on. 

In~(\ref{element-2}), we write out the generalized Laguerre
polynomials as polynomials 
in their respective arguments. 
$L^k_n(u^2/z)$ 
yields a sum of numerical coefficients times
$u^{2r}z^{-r}$, 
$ L^{k'}_{n'} \biglb(z {(v')}^2\bigrb)$ 
yields
${(v')}^{2r'}z^{r'}$, 
$L^k_{\tilde n}(u^2)$ 
yields
$u^{2s}$, 
and 
$L^{k'}_{{\tilde n}'}(v^2)$ 
yields
$v^{2s'}$, 
where $r$, $r'$, $s$, and $s'$ 
range over integers satisfying 
$0\leq r \leq n$, 
$0\leq r' \leq n'$, 
$0\leq s \leq {\tilde n}$, 
and 
$0\leq s' \leq {\tilde n}'$. 
(\ref{element-2}) equals therefore a sum over 
$r$, $r'$, $s$, and $s'$ of 
numerical coefficients times 
\begin{eqnarray}
\frac{z^{r' - r + (k'-k)/2}}
{\mu}
&&
\int 
du \, dv \, dv'
\> 
u^{2r + 2s + 2k +1}
v^{2s' + k'+1}
{(v')}^{2r' + k'+1}
J_{k'}(v v'/\mu) 
\nonumber
\\
&&
\times
\exp \left\{
- \casehalf 
\left[ 1 + (1/z) - i \mu \right] u^2 
- \casehalf 
\left[ 1  - (i/\mu) \right] v^2 
- \casehalf 
\left[ z  - (i/\mu) \right] {(v')}^2 
\right\}
\ \ .
\label{element2-expanded}
\end{eqnarray}

In~(\ref{element2-expanded}), we perform first the elementary integral 
over~$u$. We then perform the integral over $v$ using 
(6.631.10) in \cite{Grad-Rhyz}: the result involves the generalized 
Laguerre polynomial 
$L_{s'}^{k'}$ of argument 
${(v')}^2 /[2\mu(\mu-i)]$, and we expand this polynomial as a
sum of numerical coefficients times
$\left\{ {(v')}^2
  /[\mu(\mu-i)]\right\}^{s''}$, where 
$s''$ 
ranges over integers satisfying 
$0\leq s'' \leq s'$. The remaining integral over 
$v'$ is
elementary. Note that 
these integrals over $u$, $v$, and $v'$ 
converge in absolute value. 
Collecting, we find that (\ref{element-2}) is a sum 
over 
$r$, $r'$, $s$, $s'$, and $s''$ of 
numerical coefficients times 
\begin{equation}
\mu^{s' - s''}
{(1 + i\mu)}^{r'-s'}
z^{1 + r' +s  + (k'+k)/2}
{(1 + z + i\mu z)}^{-r' - s'' - k' - 1}
{(1 + z - i\mu z)}^{-r - s - k - 1}
\ \ , 
\label{element2-expanded2}
\end{equation}
where $0\le s'' \le s'$. 
An elementary analysis shows that 
(\ref{element2-expanded2}) is 
integrable in absolute value over $\{(z,\mu) \mid z>0, \mu \in
\BbbR\}$ in the measure $\int dz \, d\mu$ 
provided $r + (k + k')/2 >0$. As
$|m|+|m'|>0$ by assumption, this condition is satisfied. 
Thus (\ref{GAMtoo}) converges in absolute value. 

We note that the assumption $|m|+|m'|>0$ was only used in the final
step, in showing the 
integrability of~(\ref{element2-expanded2}). We also note that
this assumption is necessary. Taking 
$\phi_1 = \phi_{0,0;0,0}$ 
and 
$\phi_2 = \phi_{0,0;n,n}$, 
(\ref{app-rep-iwasawa}) and (\ref{element-2}) yield, using 
\cite{magnusetal-integrals}
and (6.631.4) in \cite{Grad-Rhyz}, 
\begin{equation}
\biglb(\phi_2 , U(g) \phi_1
\bigrb)_{\rm aux}
= 
4\pi^2 {(-1)}^{n}
\times 
\frac{
z 
{\left[ (1-z)^2 + \mu^2 z^2 \right]}^n}
{
{\left[ (1+z)^2 + \mu^2 z^2 \right]}^{n+1}}
\ \ ,
\label{diverg-element}
\end{equation}
and the integral of (\ref{diverg-element}) 
over the group in the Haar measure is 
unambiguously divergent.

\subsection{Both angular momenta zero}

We set 
$\phi_1 = \phi_{0,0;n,n'} + \phi_{0,0;n+1,n'+1}$ 
and 
$\phi_2 = \phi_{0,0;{\tilde n},{\tilde n}'} 
+ \phi_{0,0;{\tilde n}+1,{\tilde n}'+1}$. 
As above, it suffices to take $\theta=0$ and consider the integral
over $\lambda$ and $\mu$ in the measure 
$e^{2\lambda}\,d\lambda\,d\mu$. 

Let again $U(g)$ be as in (\ref{app-rep-iwasawa}) 
with $\theta=0$ and $\mu\neq0$. 
By~(\ref{nullanddila-action}), 
we have 
\begin{eqnarray}
\biglb(\phi_2 ,  U(g) \phi_1  \bigrb)_{\rm aux} 
&&
= 
\frac{4 \pi^2}
{i \mu}
\int 
du \, dv \, dv'
\> 
u
v
v'
J_{0}(v v'/\mu) 
\left[ 
L_{\tilde n}(u^2) 
L_{{\tilde n}'}(v^2) 
+ 
L_{{\tilde n}+1}(u^2) 
L_{{\tilde n}'+1}(v^2) 
\right] 
\nonumber
\\
\noalign{\smallskip}
&&
\times
\left[ 
L_n(u^2/z) 
L_{n'} \biglb(z {(v')}^2\bigrb) 
+ 
L_{n+1}(u^2/z) 
L_{n'+1} \biglb(z {(v')}^2\bigrb) 
\right]
\nonumber
\\
&&
\times
\exp \left\{
- \casehalf 
\left[ 1 + (1/z) - i \mu \right] u^2 
- \casehalf 
\left[ 1  - (i/\mu) \right] v^2 
- \casehalf 
\left[ z  - (i/\mu) \right] {(v')}^2 
\right\}
\ \ . 
\label{zamoelement-2}
\end{eqnarray}

In~(\ref{zamoelement-2}), we write out $L_n(u^2/z)$ and
$L_{n+1}(u^2/z)$ as polynomials in their arguments.  The previous analysis
[the integrability of (\ref{element2-expanded2}) 
for $k=k'=0$ provided $r>0$] shows that the nonconstant terms give
integrable contributions. Consider therefore the expression where 
$L_n(u^2/z)$ and
$L_{n+1}(u^2/z)$ in (\ref{zamoelement-2}) are each replaced by their
constant term~$1$. We perform the integrals over 
$u$ and $v$ using 
(7.414.6) and (7.421.1) in \cite{Grad-Rhyz}, obtaining a sum of
numerical constants times 
\begin{eqnarray}
\frac{z 
{(1 - z - i\mu z)}^{p} 
{(1 -i\mu)}^{p'} }
{{(1 + z - i\mu z)}^{p+1} 
{(1 + i\mu)}^{p'+1} }
\int 
&&
dv' \, v' 
L_{p'} \left( \frac{{(v')}^2}{1+\mu^2} \right)
\exp 
\left\{
- \frac{1}{2} 
\left[
z + \frac{1}{(1+i\mu)} 
\right]
{(v')}^2 
\right\}
\nonumber
\\
&&
\times
\left[ 
L_{n'} \biglb(z {(v')}^2\bigrb) 
+ 
L_{n'+1} \biglb(z {(v')}^2\bigrb) 
\right]
\ \ , 
\label{zamoelement-2-expanded}
\end{eqnarray}
where 
$(p,p')  = ({{\tilde n}},{{\tilde n}'})$ or 
$(p,p')  = ({{\tilde n}}+1,{{\tilde n}'}+1)$. 
In~(\ref{zamoelement-2-expanded}), we write out
$L_{p'}$ as a sum of numerical coefficients times
${(v')}^{2s}{(1+\mu^2)}^{-s}$, 
where $s$ ranges over integers
satisfying 
$0\le s \le {p'}$. 
We then perform the remaining integral by changing the
integration variable from $v'$ to $x:=z{(v')}^2$ 
and using the formula 
\begin{equation}
\int_0^\infty dx \, x^s 
\left[ L_{n'}(x) + L_{n'+1}(x) \right] 
\exp\left[ - \casehalf \left( 1 + a^{-1} \right)x \right]
= 
\frac{a^{s+1} P_{n',s}(a)}{{(1+a)}^{n'+s+2}}
\ \ ,
\label{auxint-formula}
\end{equation}
where $P_{n',s}$ is a polynomial 
(whose precise numerical coefficients will not be 
needed) of order~$n'+s$. 
The validity of (\ref{auxint-formula}) for $s=0$ 
follows from (7.414.7) in \cite{Grad-Rhyz}, and the validity for $s>0$ 
follows by repeated differentiation with respect to~$a^{-1}$. It then
follows by elementary analysis that (\ref{zamoelement-2-expanded})
is integrable in absolute value over $\{(z,\mu) \mid z>0, \mu \in
\BbbR\}$ in the measure $\int dz \, d\mu$. 
Thus (\ref{GAMtoo}) converges in absolute value.

\section{Evaluation of the rigging map}
\label{app:rigg-ev}

In this appendix we evaluate the map~$\eta$, given by 
(\ref{GAM}) and~(\ref{GAM-eta}), 
on the test function 
space $\Phi$ defined in section~\ref{sec:RAQus}. 

It suffices to consider test states 
$\phi$ in the set $B_0$~(\ref{Bnought-def}).
We consider separately the case where both angular momenta are
nonzero and the case where at least one angular momentum is zero.

\subsection{Both angular momenta nonzero}

Suppose $m\neq 0 \neq m'$, and consider $U(g) \phi_{m,m';n,n'}$ as a
function on $G\times\BbbR^4$, where $G = \SLtwor$ is the gauge group
and $\BbbR^4 = \left\{(\vec u, \vec v)\right\}$ is the configuration
space. By the methods of appendix \ref{app:convergence} it is
straightforward to show that 
$U(g) \phi_{m,m';n,n'}$ is integrable in absolute value over $G$
pointwise in $(\vec u, \vec v)$, and that 
${\overline\phi}U(g) \phi_{m,m';n,n'}$ is integrable in absolute value
over $G\times\BbbR^4$ for every $\phi\in\Phi_0$. 
It follows by Fubini's theorem that $\eta(\phi_{m,m';n,n'})$ can be
represented by a function on~$\BbbR^4$, acting on test states
$\phi\in\Phi$ by~(\ref{dual-action}): we have 
\begin{equation}
\eta(\phi_{m,m';n,n'}) = \overline{\chi_{m,m';n,n'}}
\ \ ,
\label{chi-conjugate}
\end{equation}
where 
\begin{equation}
\chi_{m,m';n,n'} := \int_G dg \, U(g) \phi_{m,m';n,n'}
\ \ ,
\label{chi-def}
\end{equation}
and the integral in (\ref{chi-def}) is evaluated pointwise on~$\BbbR^4$. 
We shall now evaluate~(\ref{chi-def}). 

We write $U(g)$ in the Iwasawa 
decomposition~(\ref{app-rep-iwasawa}) and write 
$z := e^{2\lambda}$. 
For $\mu\ne0$, we obtain 
\begin{eqnarray}
U(g) \phi_{m,m';n,n'}
= &&
\frac{
{[{\rm sgn}( m')]}^{m'}
\, 
e^{i(m\alpha + m'\beta)}
z^{(k'-k)/2}
e^{i\theta(k' - k + 2n' - 2n)}
}
{{(i)}^{m'+1} \mu}
\nonumber
\\
&&
\times
\int_0^\infty dv' \, 
u^k {(v')}^{k'+1}
J_{k'}(v v'/\mu) 
L^k_n(u^2/z) 
L^{k'}_{n'} \biglb(z {(v')}^2\bigrb) 
\nonumber
\\
&&
\qquad\qquad
\times
\exp\left[
- \frac{1}{2} 
\left(
\frac{u^2}{z} 
+ 
z {(v')}^2
\right) 
+ \frac{i}{2} 
\left( \mu u^2 + \frac{ v^2 + {(v')}^2}{\mu} \right)
\right]
\ \ ,
\label{Uphi-explicit}
\end{eqnarray}
where $k:=|m|$ and $k':=|m'|$, and by assumption 
$k\ge1$ and
$k'\ge1$. As in appendix~\ref{app:convergence}, the 
Bessel function $J_{k'}(v v'/\mu)$ has emerged from
performing the angular part of the $d^2 \vec v'$ 
integral in~(\ref{null-action}). The integral in 
(\ref{Uphi-explicit}) could be performed 
in terms of a generalized Laguerre polynomial
using (7.421.4) in \cite{Grad-Rhyz}, but for us it will be more
convenient to proceed directly with~(\ref{Uphi-explicit}). 

We now integrate (\ref{Uphi-explicit}) in the Haar measure 
$dg = e^{2\lambda}\,d\lambda\,d\mu\,d\theta 
= \casehalf dz\,d\mu\,d\theta$. By the above discussion,
this integral converges in absolute value.
We may assume $u>0$ and
$v>0$. The integral over
$\theta$ yields the factor 
$2\pi \delta_{k+2n, k'+2n'}$. 
In the remaining expression we first 
change the variable in the integral in (\ref{Uphi-explicit})
from $v'$ to
$x:=z{(v')}^2$, and we then change the variables in the outer integral 
$\int dz\,d\mu$ 
to $y := u^2/z$ and 
$p := u^2 \mu$. We obtain 
\begin{eqnarray}
\chi_{m,m';n,n'} 
&&= 
\frac{
\pi {[{\rm sgn}( m')]}^{m'}
\delta_{k+2n, k'+2n'}
\, 
e^{i(m\alpha + m'\beta)}
}
{2 {(i)}^{m'+1}}
\int_0^\infty dy \, y^{(k/2) - 1} L^k_n(y) e^{-y/2}
\nonumber 
\\
\times
&&
\int_{-\infty}^\infty \frac{dp}{p}
\int_0^\infty dx \,  x^{k'/2}
J_{k'} \left( \frac{uv \sqrt{xy}}{p} \right)
L^{k'}_{n'} (x) 
\, 
\exp\left[ - \frac{x}{2} + 
\frac{i}{2} \left( p + \frac{u^2 v^2 + xy}{p} \right) \right]
\ \ . 
\label{chi-tripleint}
\end{eqnarray}
We then interchange the order of the $\int dx$ and $\int dp$
integrals in~(\ref{chi-tripleint}), 
justified by the absolute convergence of 
the double integral $\int dx \, dp$. Performing
the $\int dp$ integral by (the absolutely convergent analytic
continuation of) (6.635.3) in \cite{Grad-Rhyz}, we obtain 
\begin{eqnarray}
\chi_{m,m';n,n'} 
=&& \pi^2 \delta_{k+2n, k'+2n'}
e^{i(m\alpha + m'\beta)}
J_{k'}(uv)
\nonumber
\\
&& \ 
\times
\int_0^\infty dy \, 
y^{(k/2) - 1} L^k_n(y) e^{-y/2}
\int_0^\infty dx \, 
x^{k'/2} J_{k'} \left(\sqrt{xy}\right)
L^{k'}_{n'}(x) e^{-x/2}
\nonumber
\\
=&&
2\pi^2 {(-1)}^{n'} \delta_{k+2n, k'+2n'}
e^{i(m\alpha + m'\beta)}
J_{k'}(uv)
\int_0^\infty dy \, 
y^{(k+k')/2 - 1} L^k_n(y) L^{k'}_{n'}(y) e^{-y}
\ \ ,
\label{chiw-lastint}
\end{eqnarray}
where in the last step we have evaluated the $\int dx$ integral using 
\cite{magnusetal-integrals}.  

Consider the remaining integral in~(\ref{chiw-lastint}). 
Suppose $k' \ge k$. 
Because of the factor~$\delta_{k+2n, k'+2n'}$, it 
suffices to consider
$k' = k + 2s$ and $n = n'+s$ for some nonnegative integer~$s$. 
We thus need to evaluate 
\begin{equation}
\int_0^\infty
dy \, 
y^{k + s - 1} 
L^k_{n'+s}(y)
L^{k+2s}_{n'}(y)
e^{-y}
\ \ .
\label{lag-powerint}
\end{equation}
Expanding 
$L^{k+2s}_{n'}(y)$ 
in (\ref{lag-powerint})
as a polynomial
in $y$ yields integrals of the form 
\begin{equation}
\int_0^\infty
dy \, 
y^{k+q} 
L^k_{n'+s}(y)
e^{-y}
\ \ ,
\label{lag-powerexp}
\end{equation}
where $s-1 \leq q \leq s + n' - 1$. The orthogonality of the
generalized Laguerre polynomials \cite{magnusetal-recurrence} 
implies that (\ref{lag-powerexp}) vanishes for
$0\leq q < n'+s$. When $s>0$, $q$~is always in this range, and
(\ref{lag-powerint}) thus vanishes. 
When $s=0$, 
the only value of $q$ not in this range is $q=-1$, which comes from 
the constant term of the
expanded $L^{k}_{n'}(y)$ in~(\ref{lag-powerint}): 
using \cite{magnusetal-definition}, (7.414.7) 
in \cite{Grad-Rhyz}, and (15.1.40) in \cite{abra-stegun}, 
we then find that 
(\ref{lag-powerint}) for $s=0$ is equal to $(n'+k)!/[k \,
(n')!]$.
Finally, the case $k' < k$ reduces to the case already considered by
interchange of the primed and unprimed indices, and we find that 
(\ref{chiw-lastint}) vanishes. 

Expressing the result in terms of the original
indices, we have 
\begin{equation}
\chi_{m,m';n,n'} 
= 
2\pi^2 {(-1)}^{n} 
{[{\rm sgn}( m)]}^{m}
\delta_{|m|, |m'|} \, \delta_{n, n'}
\, 
\frac{(n+|m|)!}{|m| \, n!}
\, 
J_m(uv) 
\, 
e^{i(m\alpha + m'\beta)}
\ \ .
\end{equation}
The result (\ref{aver-result-nonzero}) then follows from
(\ref{chi-conjugate}) and~(\ref{eta-imagebasis}).

\subsection{At least one angular momentum zero}

What remains is to evaluate the map $\eta$ for 
$\phi_{0,m';n,n'}$ with $m'\ne0$, 
$\phi_{m,0;n,n'}$  with $m\ne0$, 
and 
$\phi_{0,0;n,n'} + \phi_{0,0;n+1,n'+1}$. 
We shall show that $\eta$ vanishes on these states. 

A~direct analysis along the above lines would run into a 
technical difficulty in that not all the
analogous multiple integrals now converge
in absolute value. 
It is however suggestive to note that the 
integrals are still conditionally convergent, 
and starting from the counterpart of (\ref{chi-def}) 
and formally interchanging the integrations 
as above yields the
result zero. For 
$\phi_{0,m';n,n'}$ and 
$\phi_{m,0;n,n'}$, (\ref{chi-def})~yields the zero
function and hence the zero vector in~$\Phi^*$. 
For $\phi_{0,0;n,n'} + \phi_{0,0;n+1,n'+1}$, 
the counterpart of 
(\ref{chi-def}) yields a function proportional 
to~$J_0(uv)$, which clearly solves the quantum constraints, 
but the dual action (\ref{dual-action}) of $J_0(uv)$ on
every vector in $\Phi_0$ vanishes [by the extension of 
(\ref{fs-on-phis-nonzero}) to $m=0$], 
and as an element of $\Phi^*$
$J_0(uv)$ is thus identical to the zero vector. 
We now  
show that the result zero is indeed the correct one. 

Consider first 
\begin{equation}
\eta(\phi_{0,m';p,p'})[\phi] 
= 
{(\phi_{0,m';p,p'} , \phi )}_{\rm ga}
\ \ , 
\label{silverleg}
\end{equation}
where $m'\ne0$ and $\phi\in B_0$~(\ref{Bnought-def}). 
As noted in appendix~\ref{app:convergence}, it suffices to
consider $\phi = \phi_{0,m';n,n'}$. Using the Iwasawa
decomposition 
(\ref{app-rep-iwasawa}) in~(\ref{GAM}), 
the integral over $\theta$ shows that we can set 
$|m'| = 2(n - n')$, and a similar reasoning with 
$U(g)$ in (\ref{GAM}) conjugated to act on the first
argument shows that we can set 
$|m'| = 2(p - p')$. It therefore suffices to consider 
${(\phi_{0,\pm2s;p+s,p} , \phi_{0,\pm2s;n+s,n})}_{\rm ga}$  with 
$s\ge1$. 

Consider thus 
${(\phi_{0,2s;p+s,p} , \phi_{0,2s;n+s,n})}_{\rm ga}$ with 
$s\ge1$. 
We recall that the operators 
${\widehat\tau}_\pm^\eta$ 
(\ref{polartaus}) are in~$\Aobs$ and the adjoint of 
${\widehat\tau}_\pm^\eta$ in $\Haux$ is~${\widehat\tau}_\mp^\eta$. 
Using properties of the generalized Laguerre polynomials 
\cite{magnusetal-recurrence} we find 
\begin{mathletters}
\begin{eqnarray}
&&
{\widehat\tau}_-^- \phi_{1,2s-1;n+s-1,n}
= 
- (n+s)
\left(
\phi_{0,2s;n+s-1,n-1}
+ 
\phi_{0,2s;n+s,n}
\right)
\ \ ,
\\
&&
{\widehat\tau}_+^- \phi_{0,2s;p+s,p}
= 
-
(p+1)
\phi_{1,2s-1;p+s,p+1}
+ 
(p+2s)
\phi_{1,2s-1;p+s-1,p}
\ \ ,
\end{eqnarray}
\end{mathletters}
where 
$\phi_{0,2s;n+s-1,n-1}$ for $n=0$ is understood as the zero vector. 
We therefore have 
\begin{eqnarray}
(n+ && s) 
\left[
{(\phi_{0,2s;p+s,p} , 
\phi_{0,2s;n+s-1,n-1}
)}_{\rm ga}
+ 
{(\phi_{0,2s;p+s,p} , 
\phi_{0,2s;n+s,n}
)}_{\rm ga}
\right]
\nonumber
\\
= &&
- 
{(\phi_{0,2s;p+s,p} , 
{\widehat\tau}_-^- \phi_{1,2s-1;n+s-1,n}
)}_{\rm ga}
\nonumber
\\
= &&
- 
{( {\widehat\tau}_+^- \phi_{0,2s;p+s,p} , 
\phi_{1,2s-1;n+s-1,n}
)}_{\rm ga}
\nonumber
\\
= &&
(p+1)
{( \phi_{1,2s-1;p+s,p+1} , 
\phi_{1,2s-1;n+s-1,n}
)}_{\rm ga}
+ 
(p+2s)
{( 
\phi_{1,2s-1;p+s-1,p} , 
\phi_{1,2s-1;n+s-1,n}
)}_{\rm ga}
\nonumber
\\
= && 0
\ \ ,
\label{chain}
\end{eqnarray}
where the last equality follows from (\ref{aver-result-nonzero}) in
the index range where (\ref{aver-result-nonzero}) has already been
verified. 
By induction in~$n$, 
(\ref{chain})~implies ${(\phi_{0,2s;p+s,p} , 
\phi_{0,2s;n+s,n}
)}_{\rm ga}=0$. 
An analogous argument shows ${(\phi_{0,-2s;p+s,p} , 
\phi_{0,-2s;n+s,n}
)}_{\rm ga}=0$. 

Thus $\eta(\phi_{0,m';p,p'}) = 0$ for
$m'\ne0$. A~similar argument shows that 
$\eta(\phi_{m,0;p,p'}) = 0$ for
$m\ne0$. Finally, $\eta(\phi_{0,0;p,p'} + \phi_{0,0;p+1,p'+1} )=0$
follows by applying an analogous reasoning to the relations 
\cite{magnusetal-recurrence}
\begin{mathletters}
\begin{eqnarray}
&&
{\widehat\tau}_-^+ \phi_{1,1;n,n'}
= 
(n+1) (n'+1) 
\left( \phi_{0,0;n,n'} + \phi_{0,0;n+1,n'+1} \right) 
\ \ , 
\\
&&
{\widehat\tau}_+^+ 
\left(\phi_{0,0;p,p'} + \phi_{0,0;p+1,p'+1} \right) 
= 
\phi_{1,1;p-1,p'-1}
+ 2 \phi_{1,1;p,p'}
+ \phi_{1,1;p+1,p'+1}
\ \ .
\end{eqnarray}
\end{mathletters}

\end{document}